\documentclass[aps,prl,reprint,superscriptaddress]{revtex4-2}
\usepackage{amsmath,amsfonts,amssymb,bm,hyperref,graphicx,subfigure,orcidlink}
\usepackage[final]{changes}
\usepackage[notrig]{physics}
\usepackage{CJKutf8}
\usepackage{xcolor}

\UseRawInputEncoding
\newcommand{\ud}{\mathrm{d}}

\newcommand{\ue}{\mathrm{e}}

\begin{document}

\begin{CJK*}{UTF8}{gbsn} 

\title{Promoting Fluctuation Theorems into Covariant Forms}

\author{Ji-Hui Pei (裴继辉)\orcidlink{0000-0002-3466-4791}}
\affiliation{School of Physics, Peking University, Beijing, 100871, China}
\affiliation{Department of Physics and Astronomy, KU Leuven, 3000, Belgium}
\author{Jin-Fu Chen (陈劲夫)\orcidlink{0000-0002-7207-969X}}
\email{jinfuchen@lorentz.leidenuniv.nl}
\affiliation{School of Physics, Peking University, Beijing, 100871, China}
\affiliation{Instituut-Lorentz, Universiteit Leiden, P.O. Box 9506, 2300 RA Leiden, The Netherlands}
\author{H. T. Quan (全海涛)\orcidlink{0000-0002-4130-2924}}
\email{htquan@pku.edu.cn}
\affiliation{School of Physics, Peking University, Beijing, 100871, China}
\affiliation{Collaborative Innovation Center of Quantum Matter, Beijing 100871,
China}
\affiliation{Frontiers Science Center for Nano-optoelectronics, Peking University,
Beijing, 100871, China}

\begin{abstract}
    The principle of covariance, a cornerstone of modern physics, asserts the equivalence of all inertial frames of reference.
    Fluctuation theorems, as extensions of the second law of thermodynamics, 
    establish universal connections between irreversibility and fluctuation in terms of stochastic thermodynamic quantities. 
    However, these relations typically assume that both the thermodynamic system and the heat bath are at rest with respect to the observer, thereby failing to satisfy the principle of covariance. In this study, by introducing covariant work and heat that incorporate both energy-related and momentum-related components, we promote fluctuation theorems into covariant forms applicable to moving thermodynamic systems and moving heat baths. 
    We illustrate this framework with two examples: the work statistics of a relativistic stochastic field and the heat statistics of a relativistic Brownian motion.
    Although our study is carried out in the context of special relativity, the results can be extended to the nonrelativistic limit. 
    Our work combines the principle of covariance and fluctuation theorems into a coherent framework and may have applications in the study of thermodynamics relevant to cosmic microwave background as well as 
    the radiative heat transfer and noncontact friction between relatively moving bodies. 
\end{abstract}
\maketitle
\end{CJK*}
\textit{Introduction--}
The second law of thermodynamics is a statement about irreversibility. 
Many macroscopic phenomena occur in one direction, rarely seen in reverse. 
When examining systems at the mesoscopic scale, irreversibility is accompanied by fluctuations, which are more precisely captured by fluctuation theorems. 
Fluctuation theorems \cite{Evans1993,Gallavotti1995,Jarzynski1997,Crooks1998,Crooks1999,Seifert2005,Jarzynski2011}, 
one of the central achievements in stochastic thermodynamics \cite{Seifert2012,Peliti2021,Shiraishi2023}, 
relate the probability ratio between forward and backward random trajectories to the exponential of thermodynamic quantities associated with those trajectories. 
At the ensemble average level, the second law of thermodynamics is a corollary of fluctuation theorems. 

Stochastic thermodynamics and fluctuation theorems have been studied in various systems, including those where relativistic effects are prominent. 
Early works on Brownian motion for relativistic particles and relativistic kinetic theory \cite{Debbasch1997,Dunkel2009,PhysRevE.72.036106,Hakim2011,Vereshchagin2017} 
spawned discussions on nonequilibrium phenomena in relativistic systems. 
Most recently, attempts are made to develop the framework of stochastic thermodynamics for single particle system in Minkowski spacetime \cite{Pal2020,Paraguassu2021} and 
in curved spacetime \cite{Cai2023,Cai2023a,Wang2024}. 
For specific relativistic systems, work distributions \cite{PhysRevA.99.052508,Ortega2019,Bartolotta2018,TeixidoBonfill2020} and fluctuation theorems   \cite{PhysRevA.99.052508,Cleuren2008,Pal2020,Fingerle2007,Ortega2019,TeixidoBonfill2020,Bartolotta2018,Torrieri2021,Zhang2024} have been studied in the rest reference frame.

Behind almost all the statements of fluctuation theorems, an implicit assumption is that the heat bath and the thermodynamic system are initially at rest. 
Definitions of stochastic work and heat, as well as the formulation of fluctuation theorems, have not been extended to the system and the heat bath in motion, let alone to them in relative motion. 
As a pillar of modern physics, the principle of covariance is a fundamental requirement for all physical laws. However, the existing fluctuation theorems do not satisfy the principle of covariance. Hence, a significant unsolved problem is to promote the existing fluctuation theorems into covariant forms.

In this Letter, we promote  the existing fluctuation theorems into covariant forms within the context of special relativity. 
Nonrelativistic systems are treated as a low speed limit. 
We find that the Lorentz covariance, which links space and time, necessitates a new consistent definition of the backward process across all inertial frames.  
For moving systems, in accordance with the energy-momentum 4-vector, 
the concepts of work and heat must be generalized to 4-vector form. 
Consequently, the refined statements of the second law of thermodynamics and fluctuation theorems must account for momentum-related quantities. 
When multiple systems or heat baths initially move relative to each other, existing fluctuation theorems fail, but our covariant theorems remain valid, extending the study of irreversibility and fluctuations to broader contexts.

\textit{Covariant fluctuation theorems--}
Fluctuation theorems are universal relations in thermodynamics, independent of the underlying dynamics of the system. The detailed fluctuation theorems  relate the probability ratio between forward and backward trajectories, $\Pr(\omega)/\tilde\Pr(\tilde \omega)$, to thermodynamic quantities, and they can be expressed in the following form:
\begin{equation}\label{ft_e}
    \frac{\Pr(\omega)}{\tilde\Pr(\tilde \omega)}=\exp(\text{thermodynamic quantity}).
\end{equation}
Before proceeding to covariant fluctuation theorems, we need to specify the backward process and find the corresponding thermodynamic quantity in the context of special relativity.


We adopt the four-dimensional spacetime notation $x=(x^0=t,x^1,x^2,x^3)$ and the Einstein summation rule with metric $\eta_{\mu\nu}=\operatorname{diag}(1,-1,-1,-1)$. 
Roman scripts $i,j=1,2,3$ are used for spatial components. 
Let us consider a general thermodynamic process over a finite{\color{blue}-}time period (precisely speaking, a process between two space-like hypersurfaces). 
The system is driven by a spacetime-dependent external control field $h(x)$, so that 
the Lagrangian (or Hamiltonian) of the system explicitly depends on $h(x)$. 
Meanwhile, the system is in contact with a heat bath at the inverse temperature $\beta$. 
Usually, these problems are discussed in the rest frame of the heat bath, but we can switch to different inertial reference frames where the heat bath is moving.

\begin{figure}[htp]
        \centering
        \includegraphics{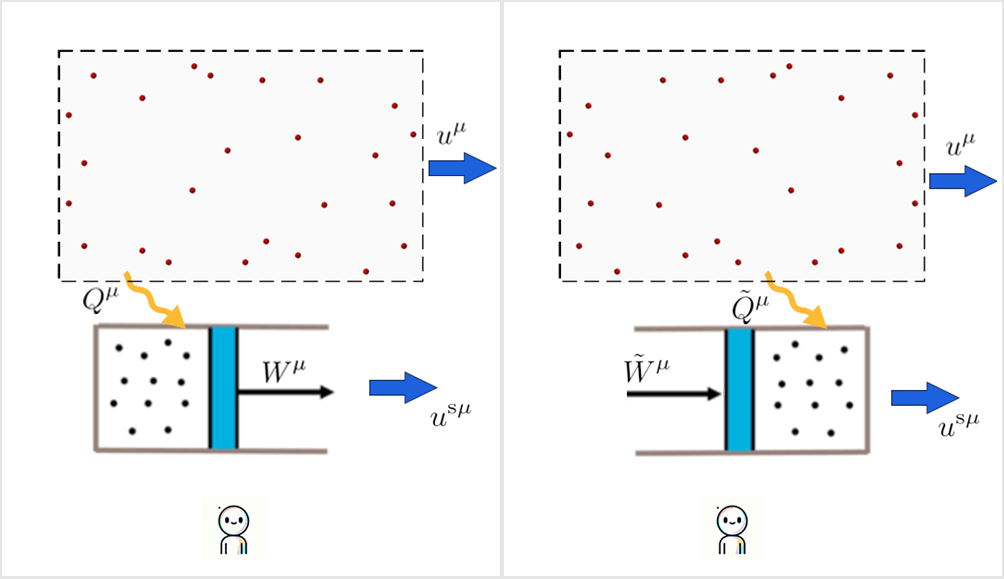}
        \caption{
        A gas-expansion process and its backward process viewed by an observer: 
        Gas molecules (depicted by black dots) are in a cylinder. 
        In the forward process (left panel), work is performed by expanding the piston, and the system exchanges heat
        with the heat bath (represented by red dots). 
        The heat bath is moving at the four-velocity $u^\mu$, and the initial 4-velocity of the system is $u^{\mathrm{s}\mu}$.
        In the backward process (right panel), the piston is compressed, 
       but the 4-velocities do not change.
        }\label{gas}
    \end{figure}

The backward process is constructed by inverting both the space and the time of the driving field, $h(x)\rightarrow h(-x)$. 
This differs slightly from the usual definition that only reverses time, in order to ensure that the backward process is frame-independent. 
Note that as an even variable under the spacetime reversal, the velocity of the heat bath does not change. 
The initial distribution of the backward process will be specified later. 
Meanwhile, we define a backward trajectory $\tilde \omega$ in the backward process as 
the spacetime reversal of the forward trajectory $\omega$ in the forward process. 
The cartoon in Fig. \ref{gas} schematically illustrates a gas-expansion process and its backward process viewed by an observer (the system and the heat bath are moving at different velocities).

We continue to identify stochastic thermodynamic quantities. 
Energy and momentum form a 4-vector under the Lorentz transformation. 
The importance of 4-momentum is firstly reflected in the canonical equilibrium distribution for a moving object, $\exp(-\beta_\mu^\mathrm{s} P^\mu)/Z$, where $P^\mu$ is the 4-momentum, and $Z$ is the partition function \cite{Kampen1968,Nakamura2009,Wu2009,Lee2017,Derakhshani2019,Hakim2011}. 
Here, the inverse temperature 4-vector of the system $\beta_\mu^\mathrm{s}=\beta^\mathrm{s} u_\mu^\mathrm{s}$ is defined as the rest inverse temperature $\beta^\mathrm{s}$ (inverse temperature observed in the reference frame where the equilibrium system is at rest on average) times the 4-velocity $u_\mu^\mathrm{s}=\eta_{\mu\nu} u^{\mathrm{s}\nu}$. 
We extend the stochastic work and heat \cite{Jarzynski2011,Seifert2012,Peliti2021,Shiraishi2023} to stochastic work 4-vector $W^\mu$ and heat 4-vector $Q^\mu$, 
corresponding to the 4-momentum change due to external driving and the 4-momentum exchange with the heat bath, respectively. 
These are stochastic generalizations of those discussed in \added{the} van Kampen formulation \cite{Kampen1968,Israel1986} and 
defined as functionals of individual stochastic trajectories. 
Furthermore, we define the stochastic entropy production of a trajectory $\omega$ as:
\begin{equation}\label{EP}
    \Sigma[\omega] = -\beta_\nu Q^\nu[\omega] +\Delta S[\omega] .
\end{equation}
Here, $\beta_\nu=\beta \eta_{\mu\nu}u^\mu$ is the inverse temperature 4-vector of the heat bath, with $\beta$ the rest inverse temperature, and $u^\mu$ the 4-velocity of the heat bath;
$\Delta S [\omega]=\ln \mathcal P_{\mathrm{ini}}(\omega_{\mathrm{ini}})-\ln\mathcal P_{\mathrm{fin}}(\omega_{\mathrm{fin}}) $ is the change of trajectory entropy (Boltzmann constant $k_\mathrm B=1$)\cite{Seifert2005};  
$\omega_{\mathrm{ini}}$ and $\omega_{\mathrm{fin}}$ are the initial and the final positions of trajectory; $\mathcal P_{\mathrm{ini}}$ and $\mathcal P_{\mathrm{fin}}$ are the initial and the final distribution functions. 
The above entropy production is a Lorentz scalar. 

We remark that there are several different formulations of relativistic thermodynamics \cite{Yuen1970,Dunkel2009,K.Nakamura2012}, one of which, the van Kampen formulation realizes a covariant theory; please see supplementary material \cite{supplementary} for an introduction of this formulation and its relation to other formulations. Our current theoretical framework can be viewed as a combination of \added{the} van Kampen formulation and stochastic thermodynamics. 

Two ingredients are at the heart of all kinds of fluctuation theorems: initial Gibbs distribution and micro-reversibility \cite{Campisi2011,Chen2023}. 
In the supplementary material \cite{supplementary}, we show how these two ingredients, with some modifications, leads to covariant fluctuation theorems.

Different fluctuation theorems apply to different setups. 
First, the fluctuation theorem for stochastic entropy production is rather general, 
which holds true for arbitrary initial distribution of the forward process. 
The initial state of the backward process is chosen to be the spacetime reversal of the final state of the forward process. 
The fluctuation theorem for entropy production reads 
\begin{equation}\label{ft_s}
    \frac{\Pr(\omega)}{\tilde \Pr(\tilde \omega)}=\ue^{\Sigma[\omega] } .
\end{equation}
Here, $\Pr$ and $\tilde \Pr$ are the trajectory probability in the forward and backward processes, respectively. 
$\omega$ and $\tilde \omega$ are a pair of forward and backward trajectories. 
Although this fluctuation theorem for entropy production appears similar to the usual version \cite{Seifert2005}, 
the difference lies in 
the specific expression for entropy production in Eq. \eqref{EP}, 
where the momentum exchange with the heat bath must be taken into account.

Second, the covariant fluctuation theorem for work applies to the following situation. 
The initial distributions of the 
forward and the backward processes are in canonical equilibrium $\exp(-\beta_\mu P^\mu(h_{\mathrm i}))/Z_i$ and $\exp(-\beta_\mu P^\mu(\tilde h_{\mathrm i}))/\tilde Z_i$, at the inverse temperature 4-vector $\beta_\mu$, same as that of the heat bath (therefore no relative motion between the system and the heat bath), 
but with respect to their own initial driving configurations, $h_{\mathrm{i}}(x)$ and $\tilde h_{\mathrm{i}}(x)=h_{\mathrm{f}}(-x)$, respectively. 
The covariant fluctuation theorem for work reads: 
\begin{equation}\label{ft_w}
    \frac{\Pr(\omega)}{\tilde \Pr(\tilde\omega)}=\ue^{\beta_\mu W^\mu[\omega]-\beta \Delta F} .
\end{equation}
Here, $\exp(-\beta \Delta F)=Z_f(\beta)/Z_i(\beta)$ is the ratio between equilibrium partition function. 
$\Delta F$ is the free energy difference.

The heat exchange fluctuation theorem applies to pure relaxation processes without external driving. 
The initial state is in equilibrium $\exp(-\beta_\mu^\mathrm{s} P^\mu)/Z_i $ at an inverse temperature 4-vector $\beta_\mu^\mathrm{s}$ different from that $\beta_\mu$ of the heat bath. 
With no external driving, the system exchanges heat 4-vector with the heat bath and evolves toward a new equilibrium state. 
The initial state of the backward process is identical to that of the forward process, 
so there is no need to distinguish $\Pr$ and $\tilde\Pr$. 
The covariant heat exchange fluctuation theorem reads: 
\begin{equation}\label{ft_q}
    \frac{\Pr(\omega)}{\Pr(\tilde \omega)}=\ue^{-(\beta_\mu-\beta^\mathrm{s}_\mu) Q^\mu[\omega]} .
\end{equation}
The initial inverse temperature 4-vector of the system $\beta_\mu^\mathrm{s}=\beta^\mathrm{s} u_\mu^\mathrm{s}$ may be unparallel to $\beta_\mu=\beta u_\mu$ of the heat bath, 
meaning the system initially moves relative to the heat bath
\footnote{Only the initial 4-velocity or the initial inverse temperature 4-vector of the system appears in the fluctuation theorem of heat exchange \eqref{ft_q} while the 4-velocity during the process does not appear explicitly. 
}.
Equations \eqref{ft_w} and \eqref{ft_q} are covariant generalizations of detailed fluctuation theorems of work \cite{Crooks1998} and heat \cite{Jarzynski2004}, respectively.

\textit{Discussion--}
The above fluctuation theorems are at the trajectory level. 
By rearranging and integrating over all trajectories, 
integral covariant fluctuation theorems follow as 
$\ev{\exp(-\Sigma)}=1$, 
$\ev{\exp(-\beta_\mu W^\mu +\beta \Delta F)}=1$, and 
$\ev{\exp((\beta_\mu -\beta^{\mathrm{s}}_\mu)Q^\mu)}=1$. 
These theorems apply to the forward processes under respective conditions, 
without involving the backward process. 
Furthermore, integral fluctuation theorems imply several modified statements of the second law for arbitrary inertial observers: 
$\ev\Sigma=\ev{-\beta_\mu Q^\mu+\Delta S}\geq 0$ for arbitrary nonequilibrium process, 
$\ev{\beta_\mu W^\mu}\geq\beta\Delta F$ for initial equilibrium system driven out of equilibrium, 
and $\ev{(\beta_\mu-\beta^{\mathrm{s}}_\mu) Q^\mu}\leq 0$ for heat exchange between the system and the heat bath. 
Accordingly, when considering moving systems or baths, all the original statements about the second law of thermodynamics 
must be modified to include the momentum effect. 

For the case that the system is initially at rest relative to the heat bath, 
the quantities $\beta u_\mu W^\mu$ and $(\beta-\beta_s)u_\mu Q^\mu$ will appear in the fluctuation theorems of work and heat exchange. 
Here, $u_\mu W^\mu=W^0_{\mathrm{rest}}$ and $u_\mu Q^\mu=Q^0_{\mathrm{rest}}$ are conventional (energy-related) thermodynamic quantities observed in the rest frame of the heat bath. 
This indicates that the rest frame of the heat bath is special, where only the energy-related term does not vanish, and the usual fluctuation theorems for work and heat exchange are recovered. 
Nevertheless, for moving observers to construct $u_\mu W^\mu$ and $u_\mu Q^\mu$, 
they usually need to measure all four components. 

On the other hand, if initially the system is moving relative to the heat bath, 
the heat exchange fluctuation theorem \eqref{ft_q} cannot be rewritten into a rest-frame version \cite{Jarzynski2004}, as it is impossible to find a reference frame where both the system and the bath are at rest. 
Examples include near-field radiative heat transfer and noncontact friction between moving bodies \cite{Volokitin2007,Volokitin2008} and the earth's velocity relative to the cosmic microwave background \cite{Peebles1968,Kogut1993,Ford2013,Planckcollaboration2014,Derakhshani2019}.
Our formula shows that $(\beta_\mu -\beta^{\mathrm{s}}_\mu)Q^\mu$ is the key quantity associated with irreversibility. 

The nonrelativistic limit of fluctuation theorems is nontrivial as well. 
When the 3-velocity of the heat bath is relatively small, $v\ll c$, the relativistic dynamics recovers Newtonian dynamics, 
and the Lorentz transformation is reduced to the Galilean transformation. 
Specifically, the transformation rule for work (the same for heat) is 
$W^0-\sum_{i=1}^3 v^i W^i=W_{\mathrm{rest}}^0+O(v^2/c^2)$. 
The momentum-related part appears in the leading order. 
For nonrelativistic systems, the covariant fluctuation theorems for work and heat exchange are reduced to the following formulas: 
\begin{align}
    \frac{\Pr(\omega)}{\tilde \Pr(\tilde\omega)}&=\ue^{\beta(W^0-\sum_{i=1}^3 v^i W^i)-\beta \Delta F} ,\\
    \frac{\Pr(\omega)}{\Pr(\tilde \omega)}&=\ue^{-(\beta -\beta^\mathrm{s} ) Q^0 +\sum_{i=1}^3(\beta v^i-\beta^\mathrm{s}v^{\mathrm{s}i})Q^i} . \label{nonrelativisitic}
\end{align}
Here, $v^i$ and $v^{\mathrm{s}i}$ are the 3-velocities of the heat bath and the system relative to the observer, respectively. 
Even in the nonrelativistic limit,
covariant fluctuation theorems must include the momentum-related components. 

\textit{Two examples--}
In the following we demonstrate the covariant fluctuation theorems in both field and particle systems, focusing on two examples: a driven relativistic stochastic field and relativistic Brownian motion of a charged particle.

First, we study field systems, which are important in relativity since interactions are transmitted via fields. 
Consider a classical massive scalar field $\phi$ driven by an external field $h$ and in contact with a heat bath. 
We adopt the following relativistic covariant equation of motion: 
\begin{equation}\label{c_eom}
    \partial^\mu\partial_\mu \phi+\kappa^\mu \partial_\mu\phi  + m^2\phi + \frac{\partial V(\phi)}{\partial\phi}= h+\sqrt{\frac{2\kappa}{\beta}}\xi(x) ,
\end{equation}
where $V(\phi)=g\phi^4/4+\cdots$ denotes the higher-order terms in the potential density; 
$g$ and $m$ are two parameters; 
$\beta$ is the rest inverse temperature of the heat bath;  
$\kappa^\mu=\kappa u^\mu$, with
$\kappa$ being the friction coefficient and $u^\mu$ being the four-velocity of the heat bath. 
In the rest frame of the heat bath, $\kappa u^\mu\partial_\mu\phi=\kappa \dot \phi $, 
denotes the friction from the heat bath. 
$\xi(x)$ is the standard white noise with the correlation $\ev{\xi(x)\xi(y)}=\delta^4(x-y)$. 
The amplitude $\sqrt{2\kappa/\beta}$ of the noise ensures the detailed balance condition. 
The above equation is a modified model A dynamics \cite{Hohenberg1977,Mallick2011} with inertia.

In relativity, the starting and ending time of the process are generalized to the space-like hypersurfaces $\mathcal A_-$ and $\mathcal A_+$, respectively. 
The spacetime region between these two surfaces is denoted by $\mathcal A$. 
The backward process starts from 
$\tilde{\mathcal A}_-=-\mathcal A_+$ and ends at $\tilde{\mathcal A}_+=-\mathcal A_-$. 
Here, the minus sign means multiplying every spacetime coordinate with $-1$. 
The driving field in the backward process is $\tilde{h}(x) = h(-x)$. 
Corresponding to a trajectory, $\{\phi(x)\}_{x\in\mathcal A}$ in the forward process, 
the backward trajectory in the backward process is $\tilde \phi(x)=\phi(-x),x\in-\mathcal A$.

For a field, the energy and momentum are characterized by the energy-momentum tensor 
$    T^{\mu\nu}=\frac{\partial \mathcal L}{\partial\partial_\mu\phi}\partial^\nu\phi-\eta^{\mu\nu}\mathcal L $, 
where $\mathcal L = \frac 12 \partial^\mu\phi \partial_\mu\phi -\frac 12 m^2 \phi^2 -V(\phi)+h \phi $
is the Lagrangian density. 
The energy-momentum tensor of the field is no longer a conserved current 
due to two distinct effects: the external driving $h$ and the interaction with the heat bath, which are identified as the work 4-vector and heat 4-vector during the process:
\begin{equation}\label{wq_field}
    W^\nu=\int_{\mathcal A}\ud^4x\frac{\partial T^{\mu\nu}}{\partial h}\dv{h}{x^\mu} ,\quad
    Q^\nu=\int_{\mathcal A}\ud^4x\frac{\ud T^{\mu\nu}}{\ud x^\mu}-W^\nu .
\end{equation}
The above definitions ensure that the work 4-vector comes from the time-dependent external driving, 
while the heat 4-vector comes from the 4-momentum exchange with the heat bath, in accord with the stochastic thermodynamics \cite{Sekimoto2010}.

Using path integral techniques for Eq. \eqref{c_eom},
the ratio of conditional probabilities between forward and backward trajectories satisfies the following relation \cite{supplementary}:
\begin{equation}\label{10}
    \frac{\Pr [\phi|\phi_{\mathcal A_-},\pi_{\mathcal A_-}]}{\tilde \Pr[\tilde \phi| \tilde \phi_{\tilde{\mathcal A}_-},\tilde \pi_{\tilde{\mathcal A}_-}]}=\ue^{-\beta_\mu Q^\mu[\phi]} .
\end{equation}
Here, $\pi=\partial_t \phi$ is the time derivative of $\phi$, and subscript $\mathcal A_-$ indicates the values on the initial hypersurface. 
The above probabilities are conditioned on their initial field configurations of $(\phi,\pi)$ and $(\tilde \phi,\tilde \pi)$, respectively. 
For different setups, multiplying with different initial distributions, 
Eq. \eqref{10} results in covariant fluctuation theorems 
\eqref{ft_s}-\eqref{ft_q} \cite{Seifert2012}. 
Detailed derivation can be found in the supplementary material \cite{supplementary}. 

For a special case where higher order terms $V=0$, and the driving process starts from the infinite past and ends in the infinite future, the joint distribution of work 4-vector can be calculated explicitly (see supplementary material \cite{supplementary}). This distribution satisfies the covariant Jarzynski equality $\ev{\exp(-\beta_\mu W^\mu +\beta \Delta F)}=1$.


Next, we study the particle system. 
Relativistic Brownian motion, a simple model 
for single particle moving in the presence of heat bath, attracts lots of attention \cite{Dunkel2009}. 
Here we focus on relativistic Ornstein-Uhlenbeck process, one specific form of relativistic Brownian motion, 
in which the friction is proportional to the velocity. 

Consider a particle with electric charge $q$ and mass $m$, which undergoes relativistic Ornstein-Uhlenbeck process and is driven by an external electromagnetic field with 4-potential $A^\mu(x)$. 
In the rest frame of the heat bath, the equation of motion is 
\begin{equation}\label{eq_p}
    \dv{p^i}{t}=f^i -\kappa \frac{p^i}{p^0} +\sqrt{\frac{2\kappa }{\beta}}\xi^i ,
\end{equation}
where $ f^i$ is the electromagnetic force, $\kappa$ is the friction coefficient, and 
$\xi^i$ is standard white noise with correlation $\ev{\xi^i(t)\xi^j(t^\prime)}=\delta_{ij}\delta(t-t^\prime)$. 
For a charged particle, the canonical 4-momentum is $P^\mu= p^\mu +q A^\mu$, 
where $p^\mu=m \ud x^\mu/\ud \tau$ is the kinetic momentum, and $\tau$ is the proper time. 
If the external field is stationary in the rest frame of the heat bath, $A^\mu(t,\vec x)=A^\mu(0,\vec x)$, 
the Gibbs distribution $\mathcal P^{\mathrm{eq}}\propto \exp(-\beta P^0)$ is the stationary solution of the above equation \cite{Hakim2011}. 
The corresponding covariant form is $\propto\exp(-\beta_\mu P^\mu)$ \cite{Hakim2011,Kampen1968,Nakamura2009,Wu2009,Lee2017,Derakhshani2019}. 

In a finite-time process, every worldline in the ensemble starts from space-like hypersurfaces $\mathcal A_-$ and ends at $\mathcal A_+$. 
The backward process  starts from $\tilde{\mathcal A}_-=-\mathcal A_+$ and ends at $\tilde{\mathcal A}_+=-\mathcal A_-$. 
The electromagnetic potential in the backward process is $ \tilde A^\mu(x)=A^\mu(-x)$. 
Correspondingly, the field strength $F^{\mu\nu}=\partial^\mu A^\nu-\partial^\nu A^\mu$ 
gets a minus sign, $\tilde F^{\mu\nu}(x)=-F^{\mu\nu}(-x)$
\footnote{The minus sign is for the electromagnetic strength but not for the electromagnetic potential because the potential appears directly in the 4-momentum}.
The backward trajectory in the backward process is obtained by reversing the spacetime coordinate of the corresponding worldline, 
$ \tilde x^\mu(s)= -x^\mu(-s)$, while keeping the 4-momentum unchanged. 

In this thermodynamic process, the heat 4-vector is identified as the 4-momentum change associated with the friction and noise. 
For every trajectory, we have 
$Q^0=\int_{t_{\mathcal A_-}}^{t_{\mathcal A_+}}\sum_i(-\kappa v^i+\sqrt{2\kappa/\beta}\xi^i)v^i\ud t$ and $Q^i=\int_{t_{\mathcal A_-}}^{t_{\mathcal A_+}}(-\kappa v^i+\sqrt{2\kappa/\beta}\xi^i)\ud t$ in the rest frame of the heat bath. 
Using the equation of motion \eqref{eq_p}, the above definitions of heat can be converted into a covariant expression, 
\begin{equation}
    Q^\mu=\Delta p^\mu -q\int F^{\mu\nu}\ud x_\nu .
\end{equation}
The above line integral is along the worldline $x(s)$, from $x_{\mathcal A_-}$ to $x_{\mathcal A_+}$. 
By noticing the Euler-Lagrange equation $\ud P^\mu/\ud s=\partial^\mu L$, with Lagrangian $L=m\sqrt{\frac{\ud x_\mu}{\ud s}\frac{\ud x^\mu}{\ud s}} +qA_\mu(x)\frac{\ud x^\mu}{\ud s}$, the 4-work is defined as $W^\mu = \int\ud s\partial^\mu{L}$ \footnote{The Euler-Lagrange equation is only valid in the absence of the heat bath, but this definition of work can be applied in the presence of the heat bath as well because we can consider the Euler-Lagrange equation of the composite system (system + bath), and the driving does not appear in the Lagrangian of the bath. }, which leads to 
\begin{equation}
    W^\mu=q\int \partial^\mu A^\nu\ud x_\nu .
\end{equation}
$Q^\mu$ and $W^\mu$ satisfy the first law of thermodynamics, $Q^\mu +W^\mu=\Delta P^\mu$.

According to path integral techniques for diffusion processes, 
the ratio of conditional trajectory probabilities between forward and backward trajectories satisfies \cite{supplementary} 
\begin{equation}\label{14}
    \frac{\Pr[x|x_{\mathcal A_-},p_{\mathcal A_-}]}{\tilde \Pr[\tilde x|\tilde x_{\tilde{\mathcal A}_-},\tilde p_{\tilde{\mathcal A}_-}]}=\ue^{-\beta_\mu Q^\mu[x]} .
\end{equation}
The above trajectory probabilities are conditioned on their initial position and momentum, $(x_{\mathcal A_-},p_{\mathcal A_-})$ and $(\tilde x_{\tilde{\mathcal A}_-},\tilde p_{\tilde{\mathcal A}_-})$, respectively. 
Similar 
to the procedure for field systems, the covariant fluctuation theorems \eqref{ft_s}-\eqref{ft_q} are obtained by multiplying both sides of Eq. \eqref{14}
with different initial distributions corresponding to different setups; see  supplementary material \cite{supplementary}. 



In a simple case of a pure relaxation process for a (1+1)D ultra-relativistic free particle, the joint distribution of heat 2-vector can be analytically solved \cite{Paraguassu2021,supplementary}, and it satisfies the covariant heat exchange fluctuation theorem, $\ev{\exp((\beta_\mu -\beta^{\mathrm{s}}_\mu)Q^\mu)}=1$. 


\textit{Conclusion--}
The existing fluctuation theorems are valid only when both the system and the heat bath are at rest relative to the observer and
do not satisfy the principle of covariance, a fundamental guiding principle of physical laws. We introduce the concept of covariant stochastic work 4-vector, heat 4-vector, and entropy production and then successfully promote existing fluctuation theorems into covariant forms, which are applicable to moving systems and heat baths in both relativistic and nonrelativistic systems. We demonstrate the validity of the covariant fluctuation theorems in two examples. 



By combining \added{the }van Kampen formulation and stochastic thermodynamics, our approach sheds new lights on the relativistic thermodynamics theory \cite{supplementary}.
Also, our study bridges a gap between stochastic thermodynamics and the principle of covariance. 
The covariant fluctuation theorems are significant both theoretically and practically. 
They unify fluctuation theorems for different inertial observers and show the second law should include the momentum-related quantities for moving systems and heat baths. 
Importantly, for systems in relative motion, existing fluctuation theorems fail. Only the covariant fluctuation theorems are valid. 

The study of irreversibility and fluctuations is extended  to broader contexts, for  moving systems and heat baths. 
For example, the earth is moving relative to the cosmic microwave background (CMB) radiation \cite{Peebles1968,Smoot1977,Kogut1993,Derakhshani2019,Planckcollaboration2014,Ford2013}, so the momentum-related component cannot be ignored when measuring CMB spectrum at a high precision. A precise CMB spectrum encodes
important information about the physical properties of the early universe. 
Besides, at nano-scales, near-field radiative heat transfer occurs between bodies in relative motion, accompanied by noncontact friction \cite{Volokitin2007,Volokitin2008,Stipe2001,Kuehn2006}. 
Theoretical predictions for the noncontact friction have varied widely \cite{Milton2016}.
By investigating the heat and momentum transfer via the noncontact friction, 
covariant fluctuation theorems can be used as a criterion 
to test the validity of those results.
\added{Furthermore, our work has 
potential applicability to heavy-ion physics and relativistic hydrodynamics theory when implementing thermodynamic fluctuations \cite{An2019,Bhambure2024,Mullins2023a,Mullins2023b}.}
Applications of covariant fluctuation theorems to those systems\deleted{ in relative motion} will be given in our future study.

\begin{acknowledgments}
This work is supported by the National Natural
Science Foundation of China (NSFC) under Grants No. 11825501, No. 12375028, No. 12147157.
\end{acknowledgments}

\bibliography{covariant_title_protected}

\end{document}


\begin{CJK*}{UTF8}{gbsn} 

\title{Supplemental Material: Promoting Fluctuation Theorems into Covariant Forms}

\author{Ji-Hui Pei (裴继辉)\orcidlink{0000-0002-3466-4791}}
\affiliation{School of Physics, Peking University, Beijing, 100871, China}
\affiliation{Department of Physics and Astronomy, KU Leuven, 3000, Belgium}
\author{Jin-Fu Chen (陈劲夫)\orcidlink{0000-0002-7207-969X}}
\email{jinfuchen@lorentz.leidenuniv.nl}
\affiliation{School of Physics, Peking University, Beijing, 100871, China}
\affiliation{Instituut-Lorentz, Universiteit Leiden, P.O. Box 9506, 2300 RA Leiden, The Netherlands}
\author{H. T. Quan (全海涛)\orcidlink{0000-0002-4130-2924}}
\email{htquan@pku.edu.cn}
\affiliation{School of Physics, Peking University, Beijing, 100871, China}
\affiliation{Collaborative Innovation Center of Quantum Matter, Beijing 100871,
China}
\affiliation{Frontiers Science Center for Nano-Optoelectronics, Peking University,
Beijing, 100871, China}
\maketitle
\date{\today}
\end{CJK*}
This supplemental material includes the following contents. In Sec. \ref{sec:Covariant fluctuation theorems from micro-reversibility} and Sec. \ref{sec:2}, we provide a general derivation of the covariant fluctuation theorems by considering a composite (system + bath) system. 
In Sec. \ref{sec:Proof of covariant Fluctuation theorems in two examples}, we prove covariant fluctuation theorems for two examples in the main text where the evolutions of the systems are governed by stochastic dynamics. 
In Sec. \ref{sec:work distribution for the driven field} and Sec. \ref{sec:heat distribution for relativistic OU process}, we calculate the work distribution for a field and the heat distribution for a particle\added{, respectively}. 
In Sec. \ref{formulation}, we briefly summarize formulations of relativistic thermodynamics and discuss new insights from the combination of the van Kampen formulation and stochastic thermodynamics.

\section{Covariant fluctuation theorems derived from microreversibility}\label{sec:Covariant fluctuation theorems from micro-reversibility}
We briefly illustrate how microreversibility, as well as initial canonical distributions, leads to covariant fluctuation theorems. 
We follow the procedures in Ref. \cite{Chen2023}, which derives fluctuation theorems from Newtonian dynamics. 
The difference in our study is that the time reversal is replaced by a spacetime reversal, and attention should be paid to the covariance of the framework. 

We examine a composite system consisting of a thermodynamic system and a heat bath. 
The system degrees of freedom are denoted as $\zeta$ while the heat bath degrees of freedom are denoted as $\eta$. 
$\Gamma=(\zeta,\eta)$ represents the degrees of freedom of the composite system. 
We focus on a finite-time driving process, 
which starts at the initial spacelike hypersurface $\mathcal A_-$ and ends at the final spacelike hypersuface $\mathcal A_+$. 
The initial and final phase space positions, corresponding to the values on these hypersurfaces, are denoted as $\Gamma(\mathcal A_-)$ and $\Gamma(\mathcal A_+)$. 
The trajectories of the thermodynamic system, the heat bath, and the composite system during the process are denoted as $\omega_\zeta$, $\omega_\eta$, and $\omega_\Gamma$, respectively. 

As is usually assumed in stochastic thermodynamics, we focus on the case where the system and the bath are weakly coupled. 
The 4-momentum of the system and the bath is denoted as $P^\mu_\zeta$ and $P^\mu_\eta$. 
The driving field $h(x)$ acts only on the system, meaning that $P^\mu_\eta$ does not depend on $h(x)$. 
Initially, the composite system is described by a product distribution 
\begin{equation}
    \rho_\Gamma(\zeta,\eta)=\rho_\zeta(\zeta(\mathcal A_-))\pi_\eta(\eta(\mathcal A_-)), 
\end{equation}
where the heat bath satisfies a canonical equilibrium distribution, 
\begin{equation}
    \pi_\eta(\eta(\mathcal A_-))=\exp\left[-\beta_\mu P^\mu_\eta(\eta(\mathcal A_-))\right]/Z.
\end{equation}
Here, $\beta_\mu$ denotes the inverse temperature 4-vector of the heat bath. 
For the moment, we leave the system distribution $\rho_\zeta$ unspecified. 

In the main text, we introduce the concepts of stochastic work 4-vector and stochastic heat 4-vector. 
Following basic ideas in stochastic thermodynamics, they can be generally defined in the current setup as follows. 
For a specific trajectory of the composite system, the work 4-vector is identified as the 4-momentum change of the composite system while the heat 4-vector is the 4-momentum change of the heat bath, 
\begin{align}
    &W^\mu=P^\mu_\zeta(\zeta(\mathcal A_+),h(\mathcal A_+)) +P^\mu_\eta(\eta(\mathcal A_+)) - P^\mu_\zeta(\zeta(\mathcal A_-),h(\mathcal A_-)) - P^\mu_\eta(\eta(\mathcal A_-)) ,\\
    &Q^\mu=-P_\eta^\mu(\eta(\mathcal A_+)) +  P^\mu_\eta(\eta(\mathcal A_-)) .
\end{align}
Here, the initial and the final 4-momentum is evaluated on the initial and final spacelike hypersurfaces, $\mathcal A_-$ and $\mathcal A_+$.

An equivalent definition of $W^\mu$ and $Q^\mu$ is expressed solely in terms of the system trajectory $\omega_\zeta$. 
From the Euler-Lagrange equation of the composite system, we know $\dv{x^\mu} T^{\mu\nu}=-\pdv h \mathcal L \partial^\nu h$ for a field system and $\dv{s} P^\nu=-\pdv h L \partial^\nu h$ for a particle system, where $ T^{\mu\nu}$ and $P^\nu$ are the energy-momentum tensor and 4-momentum of the composite system. 
Since the Lagrangian of the heat bath does not depend on the driving $h(x)$, $\mathcal L$ and $L$ can be replaced by the Lagrangian density $\mathcal L_\zeta$ (for field system) and the Lagrangian $L_\zeta$ (for particle system) of the thermodynamic system only. 
Thus, for field systems, we have the following definition using only the system trajectory $\omega_\zeta$,
\begin{align}
    &W^\mu[\omega_\zeta] = -\int_\mathcal{A}\ud^4 x \pdv{h}\mathcal L_\zeta \pdv{h}{x_\mu} , \label{work_system}\\
    &Q^\mu[\omega_\zeta] = P^\mu_\zeta(\zeta(\mathcal A_+),h(\mathcal A_-))- P^\mu_\zeta(\zeta(\mathcal A_-),h(\mathcal A_-))-W^\nu[\omega_\zeta] .\label{euqi_defi}
\end{align}
Here, the integral in Eq.~\eqref{work_system} is performed over the spacetime between two hypersurfaces $\mathcal A_-$ and $\mathcal A_+$. 
For field systems, we have $-\pdv{\mathcal L_\zeta}{h} \partial^\mu h=\pdv{T^{\nu\mu}_\zeta}{h}\partial_\nu h$, which shows that the above definition is equivalent to the expression of Eq. (9) in the main text. 
For particle systems, the definition is given by 
\begin{equation}\label{wq_par}
\begin{aligned}
    &W^\mu[\omega_\zeta] = -\int\ud s \pdv{h}L_\zeta \pdv{h}{x_\mu} , \\
    &Q^\mu[\omega_\zeta] = P^\mu_\zeta(\zeta(\mathcal A_+),h(\mathcal A_-))- P^\mu_\zeta(\zeta(\mathcal A_-),h(\mathcal A_-))-W^\nu[\omega_\zeta] .
\end{aligned}
\end{equation}
Here, the integral is taken along the worldline of the particle. 
The above expression \eqref{wq_par} is a generalization of Eqs. (12) and (13) of the main text. 

The key point is that $W^\mu$ and $Q^\mu$ can be determined by only the system trajectory $\omega_\zeta$, without knowing the trajectory of the heat bath. 
By the way, this statement only holds in the case that the system interacts with a single heat bath; the case of multiple heat baths \added{(within Newtonian dynamics)} is considered in Ref. \cite{Chen2023}. 

Since the dynamics of the composite system is deterministic, the trajectory $\omega_\Gamma$ is solely determined by the initial state $\Gamma(\mathcal A_-)$. 
We define the trajectory probability of the composite system conditioned on the initial state, as 
\begin{equation}\label{s7}
     \Pr[\omega_\Gamma|\Gamma(\mathcal A_-)] .
\end{equation}
This conditional probability contains delta functions, and it is nonzero only when the trajectory satisfies the deterministic dynamics. 
The joint distribution of the initial and the final state of the bath $\eta(\mathcal A_-),\eta(\mathcal A_+)$ and the trajectory of the system $\omega_\zeta$, conditioned on the initial system state $\zeta(\mathcal A_-)$, is obtained by multiplying \eqref{s7} with the bath initial distribution and \added{then }integrating out degrees of freedom of the heat bath,
\begin{equation}\label{joint_prob}
    \Pr [\omega_\zeta,\eta(\mathcal A_-),\eta(\mathcal A_+)|\zeta(\mathcal A_-)]=\int_{\eta(\mathcal A_-),\eta(\mathcal A_+)}\uD \omega_\eta \Pr[\omega_\Gamma|\Gamma(\mathcal A_-)]\pi_\eta(\eta(\mathcal A_-)), 
\end{equation}
where $\pi_\eta=\exp(-\beta_{\mu}P^\mu)/Z_\eta$ is the initial equilibrium distribution of the bath. $\int_{\eta(\mathcal A_-),\eta(\mathcal A_+)}\mathrm{D}\omega_\eta$ denotes the path integral with respect to the bath trajectory, 
which is performed under the constraint that the initial and the final points of the bath are fixed to be $\eta(\mathcal A_-)$ and $\eta(\mathcal A_+)$, and consequently, the distribution $\pi_\eta$ can be taken out of the integral. 

In the following, we discuss the backward process and microreversibility. 
Recall that in the present covariant framework, the backward process is taken as the spacetime reversal, where the driving field changes to $\tilde h(x)=h(-x)$, and the initial and the final hypersurfaces are $\tilde{\mathcal A}_-=-\mathcal A_+$ and $\tilde {\mathcal A}_+=-\mathcal A_-$. 
For every trajectory $\omega_\Gamma$ in the forward process, we introduce a spacetime-reversal trajectory $\tilde \omega_{\tilde\Gamma}$ in the backward process. 
To analyze it, we take a spacelike hypersurface $\mathcal C$ as a time slice inside the forward process. 
The spacetime-reversal trajectory satisfies $\tilde \Gamma(-\mathcal C) =\Theta\Gamma(\mathcal C)$, 
where $\Theta$ represents the operation of inverting the spacetime while keeping the momentum invariant for all particles of the system and the heat bath. 


The initial distribution in the backward process is also taken as a product distribution, 
\begin{equation}
    \tilde \rho_\zeta (\tilde \zeta(\tilde{\mathcal A}_-))\pi_\eta(\tilde \eta(\tilde{\mathcal{A}}_-)), 
\end{equation}
where the heat bath distribution $\pi_\eta$ is the same canonical equilibrium distribution as that in the forward process. 
Again, we leave $\tilde \rho_\zeta$ unspecified. 
Microreversibility implies that 
if a trajectory of the composite system $\omega_\Gamma$ satisfies the deterministic dynamics, its backward conjugate $\tilde \omega_{\tilde \Gamma}$ satisfies the backward dynamics. 
As a consequence, the conditional probability
of the spacetime-reversal trajectory in the backward process satisfies
\begin{equation}\label{microreversibility}
    \tilde \Pr[\tilde\omega_{\tilde \Gamma}|\tilde \Gamma(\tilde{\mathcal A}_-)]=\Pr[\omega_\Gamma|\Gamma(\mathcal A_-)] .
\end{equation}

In the backward process, the conditional trajectory probability
corresponding to Eq. \eqref{joint_prob} is 
\begin{equation}
    \tilde { \Pr}[\tilde\omega_{\tilde \zeta},\tilde \eta(\tilde {\mathcal A}_-),\tilde \eta(\tilde {\mathcal A}_+)|\tilde \zeta(\tilde {\mathcal A}_-)]=\int_{\tilde \eta(\tilde{\mathcal A}_-),\tilde \eta(\tilde{\mathcal A}_+)}\uD \tilde\omega_{\tilde \eta}\tilde {\Pr}[\tilde\omega_{\tilde \Gamma}|\tilde \Gamma(\tilde{\mathcal A}_-)] \pi_\eta(\tilde \eta(\tilde {\mathcal A}_-)) .
\end{equation}
Using the consequence of microreversibility, Eq. \eqref{microreversibility}, 
we find 
\begin{equation}\label{cor}
\begin{split}
    \tilde { \Pr}[\tilde \omega_{\tilde \zeta},\tilde \eta(\tilde {\mathcal A}_-),\tilde \eta(\tilde {\mathcal A}_+)|\tilde \zeta(\tilde {\mathcal A}_-)]
    &=\int_{\tilde \eta(\tilde{\mathcal A}_-),\tilde \eta(\tilde{\mathcal A}_+)}\uD \tilde\omega_{\tilde \eta} {\Pr}[\omega_\Gamma| \Gamma(\mathcal A_-)]\pi_\eta(\eta(\tilde{\mathcal A}_-))\\
    &=
    \int_{\eta(\mathcal A_-),\eta(\mathcal A_+)}\uD  \omega_\eta {\Pr}[\omega_\Gamma| \Gamma(\mathcal A_-)] \pi_\eta (\tilde \eta(\mathcal{\tilde A}_-))\\
    &=
    {\Pr}[\omega_\zeta,\eta({\mathcal A}_-),\eta( {\mathcal A}_+)| \zeta(\mathcal A_-)] \pi_\eta(\tilde \eta(\tilde {\mathcal A}_-))/\pi_\eta(\eta(\mathcal A_-)) .
\end{split}
    \end{equation}
In the second equality, we change the integral variable from $\uD \tilde\omega_{\tilde \eta}$ to $\uD \omega_\eta$. 
In the third equality, we use Eq. \eqref{joint_prob}. 
The relation between the initial point of the backward trajectory and the final point of the forward trajectory is $\tilde \eta(\tilde {\mathcal A}_-) =\Theta \eta(\mathcal A_+)$. 
If the Lagrangian of the heat bath respects the symmetry of spacetime reversal (so-called $\mathcal {PT}$-symmetry), then the 4-momentum satisfies $P^\mu(\tilde \eta(\tilde {\mathcal A}_-))=P^\mu(\Theta \eta(\mathcal A_+))=P^\mu(\eta(\mathcal A_+))$. 
The last term in Eq. \eqref{cor} is written as $ \pi_\eta(\tilde \eta(\tilde {\mathcal A}_-))/\pi_\eta(\eta(\mathcal A_-)) = \exp\{-\beta_{\mu}[P_\eta^\mu(\eta(\mathcal A_+))-P_\eta^\mu(\eta(\mathcal A_-))]\}=\exp(\beta_\mu Q^\mu)$, with $Q^\mu$ the 4-heat of the forward trajectory. 
Thus, we obtain the following relation 
\begin{equation}\label{ratio2b}
    {\Pr[\omega_\zeta,\eta(\mathcal A_-),\eta(\mathcal A_+)|\zeta(\mathcal A_-)]}=\tilde \Pr[\tilde\omega_{\tilde \zeta},\tilde \eta(\tilde {\mathcal A}_-), \tilde \eta(\tilde{\mathcal A}_+)|\tilde \zeta(\tilde {\mathcal A}_-)]\exp(-\beta_\mu Q^\mu) .
\end{equation}
We notice $\delta^4(Q^\mu-P^\mu_\eta(\eta(\mathcal A_-))+P^\mu_\eta(\eta(\mathcal A_+)))=\delta^4(\tilde Q^\mu-P^\mu_\eta(\tilde \eta(\tilde{\mathcal A}_-))+P^\mu_\eta(\tilde \eta(\tilde{\mathcal A}_+)))$, with $\tilde Q^\mu=-Q^\mu$. 
Multiplying both sides of Eq. \eqref{ratio2b} by these two delta functions and then integrating out the degrees of freedom of the heat bath, we obtain
\begin{equation}
    \frac{\Pr[\omega_\zeta,Q^\mu|\zeta(\mathcal A_-)]}{\tilde \Pr[\tilde\omega_{\tilde \zeta},-Q^\mu|\tilde \zeta(\tilde {\mathcal A}_-)]}=\exp(-\beta_\mu Q^\mu) .
\end{equation}
According to Eqs. \eqref{euqi_defi} and \eqref{wq_par}, the heat $Q^\mu = -P^\mu(\eta(\mathcal A_+))+P^\mu(\eta(\mathcal A_-))=Q^\mu[\omega_\zeta]$ can be uniquely determined by the trajectory of the system $\omega_\zeta$ in the single-heat-bath case. 
Therefore, we can get rid of the joint probability of $Q^\mu$ and obtain the final expression 
\begin{equation}\label{general_ratio}
    \frac{\Pr[\omega_\zeta|\zeta(\mathcal A_-)]}{\tilde \Pr[\tilde\omega_{\tilde \zeta}|\tilde \zeta(\mathcal A_-)]}=\exp(-\beta_\mu Q^\mu[\omega_\zeta]).
\end{equation}
The remaining derivation from Eq. \eqref{general_ratio} to various covariant fluctuation theorems (Eqs. (3)-(5) in the main text) is given in next section.

\section{From the ratio of conditional probabilities to fluctuation theorems}\label{sec:2}
We show the derivation from Eq. \eqref{general_ratio} to covariant fluctuation theorems (Eqs. (3)-(5) in the main text). 
In this section, for simplicity,  we use $\omega$ to denote the trajectory of degrees of freedom of the system; 
$\omega_-$ and $\omega_+$ represent the initial and the final configurations; 
$h_-$ and $h_+$ denote the initial and the final driving fields. 

Equation \eqref{general_ratio} is now expressed as 
\begin{equation}\label{basic}
    \frac{\Pr[\omega|\omega_-]}{\tilde \Pr[\tilde \omega|\tilde \omega_-]}=\exp(-\beta_\mu Q^\mu[\omega]) .
\end{equation}
Based on this, we further take into account the initial distribution, $\mathcal P(\omega_-)$, in different setups. 
Under the condition of the fluctuation theorem for entropy production, the initial distribution of the backward process is the spacetime reversal of the final distribution of the forward process, 
$\tilde{\mathcal P}(\tilde\omega_-)=\mathcal P(\omega_+)$. 
We multiply both sides of Eq. \eqref{basic} by $\mathcal P (\omega_-)/\tilde {\mathcal P}(\tilde\omega_-)=\exp(\Delta S)$, 
where $\Delta S=\ln \mathcal P (\omega_-)-\ln\mathcal P (\omega_+)$ is the stochastic entropy change of the system (Boltzmann constant is set to $1$). 
This leads to the fluctuation theorem for total entropy production, 
\begin{equation}
    \frac{\Pr[\omega]}{\tilde\Pr[\tilde \omega]}=\exp(\Sigma [\omega]) .
\end{equation}
Here, $\Sigma =\Delta S -\beta_\mu Q^\mu$ is the total entropy production. 
This is Eq.~(3) in the main text.

Next, we consider the covariant fluctuation theorem for work. The initial distributions of the forward and the backward processes
are in equilibrium, $\mathcal P^{\mathrm {eq}}(\omega_-)=\exp(-\beta_\mu P^\mu(\omega_-,h_-))/Z$ and $ \tilde {\mathcal P}^{\mathrm {eq}}(\tilde \omega_-)=\exp(-\beta_\mu \tilde P^\mu(\tilde\omega_-,\tilde h_-))/\tilde Z$, with respect to their own initial driving fields, $h_-(x)$ and $\tilde h_-(x)=h_+(-x)$.
We notice $\tilde P^\mu(\tilde\omega_-,\tilde h_-)-P^\mu(\omega_-,h_-)=P^\mu(\omega_+,h_+)-P^\mu(\omega_-,h_-)$ and $\tilde Z/Z=\exp(-\beta\Delta F)$.
Multiplying both sides of Eq. \eqref{basic} by $\mathcal P^{\mathrm {eq}}(\omega_-)/\tilde {\mathcal P}^{\mathrm {eq}}(\tilde\omega_{-})=\exp(\beta_\mu (P^\mu_+-P^\mu_-)-\beta\Delta F)$ and 
using the first law of thermodynamics, we obtain the fluctuation theorem for work, 
\begin{equation}
    \frac{\Pr[\omega]}{\tilde\Pr[\tilde \omega]}=\exp(\beta_\mu W^\mu [\omega]-\beta\Delta F) .
\end{equation}
This is Eq.~(4) in the main text.

For the covariant fluctuation theorem for heat exchange, 
there is no external driving.
The initial distributions of both the forward and backward processes are the same equilibrium distribution, $\mathcal P^{\mathrm {eq}}(\omega_-)=\exp(-\beta_\mu^\mathrm s P^\mu(\omega_-))/Z$ and $  {\mathcal P}^{\mathrm {eq}}(\tilde \omega_-)=\exp(-\beta_\mu^\mathrm s\tilde P^\mu(\tilde\omega_-))/ Z$,
at 4-vector inverse temperature $\beta_\mu^\mathrm s$. 
We notice that the difference of initial 4-momentum between forward and backward trajectories is the 4-heat exchanged with the bath, $\tilde P^\mu(\tilde\omega_-)-P^\mu(\omega_-)=P^\mu(\omega_+)-P^\mu(\omega_-)=Q[\omega]$.
We multiply both sides of Eq. \eqref{basic} by $\mathcal P^{\mathrm{eq}}(\omega_-)/ \mathcal P^{\mathrm{eq}}(\tilde \omega_-)=\exp(\beta_\mu^\mathrm s Q^\mu)$.
This leads to the fluctuation theorem for heat exchange, 
\begin{equation}
    \frac{\Pr[\omega]}{\Pr[\tilde \omega]}=\exp((\beta_\mu^\mathrm s-\beta_\mu) Q^\mu [\omega]),
\end{equation}
which is Eq.~(5) in the main text.


\section{Proof of covariant Fluctuation theorems in two examples}\label{sec:Proof of covariant Fluctuation theorems in two examples}

In this section, we provide alternative derivations of covariant fluctuation theorems in the two examples of the main context: stochastic relativistic scalar field and relativistic Brownian motion. 
The derivation is based on the reduced stochastic equations of motion where the degrees of freedom of the heat bath do not appear explicitly. 

\subsection{Conditional probability ratio for field}
We start with the covariant equation of motion for the scalar field in the main text, 
\begin{equation}\label{c_eom}
    \partial^\mu\partial_\mu \phi+\kappa u^\mu \partial_\mu\phi  + m^2\phi + \frac{\partial V}{\partial\phi}(\phi)= h+\sqrt{\frac{2\kappa}{\beta}}\xi(x).
\end{equation}
The direct calculation yields the following expressions of work and heat 4-vectors, 
\begin{equation}\label{wq}
    \begin{aligned}
    &W^\nu=\int_{\mathcal A}\ud^4x(-\phi \partial^\nu h) ,\\
    &Q^\nu=\int_{\mathcal A}\ud^4x\partial^\nu\phi(\partial^\mu\partial_\mu\phi+m^2\phi+\pdv{\phi}V-h) .
    \end{aligned}
\end{equation}
The probability of a trajectory conditioned on the initial configuration can be obtained by utilizing the Onsager-Machlup path integral approach to stochastic processes \cite{PhysRev.91.1505,Wio2013}, 
\begin{equation}\label{prfield}
    \Pr[\phi|\phi_{\mathcal A_-},\pi_{\mathcal A_-}]\propto \exp(-\int_{\mathcal A}\ud^4x\alpha) .
\end{equation}
Here, $\phi_{\mathcal A_-}$ and $\pi_{\mathcal A_-}=\partial_t \phi_{\mathcal A_-}$ are the field and its time derivative on the initial hypersuface $\mathcal A_-$. 
We call $\alpha$ the pseudo-Lagrangian, which is a function of $\phi$ and its spacetime derivatives, 
\begin{equation}
    \alpha = \frac{\beta}{4\kappa}[\partial^\mu\partial_\mu\phi+\kappa^\mu\partial_\mu\phi+m^2\phi+\pdv{\phi}V-h]^2 .
\end{equation}
The spacetime integral of $\alpha$ is called the pseudo-action. 

In the backward process, the driving field $\tilde h(x)=h(-x)$ undergoes a spacetime reversal. 
For the conjugate trajectory $\tilde \phi(x) =\phi(-x)$ in the backward process, the conditional probability is similarly expressed as (with the same proportionality coefficient) 
\begin{equation}
    \tilde \Pr[\tilde \phi| \tilde \phi_{\tilde{\mathcal A}_-},\tilde \pi_{\tilde{\mathcal A}_-}]\propto \exp(-\int_{\tilde {\mathcal A}}\ud^4x \tilde \alpha) .
\end{equation}
After changing the variable from $\tilde \phi(x)$ to $\phi(x)$ \replaced{and from $\tilde h(x)$ to $h(x)$}{by integrating by part}, \replaced{we find}{the pseudo-action is }
\begin{equation}\label{pactb}
    \begin{aligned}
    &\int_{\tilde{\mathcal A}}\ud^4x\tilde \alpha=\int_{\tilde{\mathcal A}}\ud^4x 
    \frac{\beta}{4\kappa}[\partial^\mu\partial_\mu\tilde\phi+\kappa^\mu\partial_\mu\tilde \phi+m^2\tilde \phi+V^\prime(\tilde \phi)-\tilde h]^2 =\int_{\mathcal A}\ud^4x 
    \frac{\beta}{4\kappa}[\partial^\mu\partial_\mu\phi-\kappa^\mu\partial_\mu\phi+m^2\phi+V^\prime(\phi)-h]^2 .
    \end{aligned}
\end{equation}
There is a minus sign for the first-order derivative term $\kappa^\mu \partial_\mu\phi$ in the pseudo-Lagrangian compared to that of the forward trajectory. 
From \replaced{Eqs. \eqref{prfield}-\eqref{pactb}}{the above expressions for the pseudo-actions of conjugate trajectories} and the expression for the heat 4-vector, the ratio between $\Pr[\phi|\phi_{\mathcal A_-}]$ and $\Pr[\tilde\phi|\tilde\phi_{\tilde{\mathcal A}_-}]$ yields 
\begin{equation}\label{ratio_field}
    \frac{\Pr [\phi|\phi_{\mathcal A_-},\pi_{\mathcal A_-}]}{\tilde \Pr[\tilde \phi| \tilde \phi_{\tilde{\mathcal A}_-},\tilde \pi_{\tilde{\mathcal A}_-}]}=\ue^{-\beta u_\mu Q^\mu[\phi]}.
\end{equation}
This is Eq.(10) of the main text. 
The remaining derivation from the above ratio to covariant fluctuation theorems is given in Sec. \ref{sec:2}. 

\subsection{Conditional probability ratio for particle}
In the example of relativistic Ornstein-Uhlenbeck process, 
we consider a charged particle moving in an electromagnetic field and interacting with a heat bath. 
In the main text, we have the \added{following }equation of motion in the rest frame of the heat bath, 
\begin{equation}\label{eq_p}
    \dv{p^i}{t}=f^i -\kappa \frac{p^i}{p^0} +\sqrt{\frac{2\kappa }{\beta}}\xi^i .
\end{equation}
Here, $\vec f=q\vec E+q\vec v\times \vec B$ is the electromagnetic force on the particle. $\vec E$ and $\vec B$ are the electric and magnetic fields, respectively. 
The path integral approach to stochastic processes \cite{PhysRev.91.1505,Wio2013} 
gives the path probability for the above equation of motion. 
For the forward trajectory $x(s)$, the path probability conditioned on the initial position and momentum $(x_{\mathcal A_-},p_{\mathcal A_-})$ is (in the rest frame of the heat bath) 
\begin{equation}
    \Pr[x|x_{\mathcal A_-},p_{\mathcal A_-}]\propto \exp[-\frac{\beta}{4\kappa}\int_{t_{\mathcal A_-}}^{t_{\mathcal A_+}}\ud t (\dv{\vec p}{t}-q\vec E-q \vec v\times \vec B +\kappa \frac{\vec p}{p^0})^2].
\end{equation}
Here, ${t_{\mathcal A_-}}$ and ${t_{\mathcal A_+}}$ are the starting and the ending time of the trajectory. 

According to the general definition in the main text, 
in the backward process, the electric and magnetic fields are $\tilde E^i(x)=-E^i(-x)$ and $\tilde B^i(x)=-B^i(-x)$, respectively.
The backward trajectory is $\tilde x^i(\tilde t)=-x^i(-t)$. 
For the backward trajectory, the conditional path probability is 
\begin{equation}
    \begin{aligned}
    \tilde \Pr[\tilde x|\tilde x_{\tilde {\mathcal A}_-},p_{\tilde {\mathcal A}_-}]&\propto\exp[-\frac{\beta}{4\kappa}\int_{\tilde t_{\tilde {\mathcal A}_-}}^{\tilde t_{\tilde {\mathcal A}_+}}\ud \tilde t(\dv{\tilde {\vec p}}{\tilde t}-q \tilde {\vec E}(\tilde {\vec x}(\tilde t),\tilde t)-q \tilde {\vec  v}\times \tilde {\vec B}(\tilde {\vec x}(\tilde t),\tilde t)+\kappa \frac{\tilde {\vec p}}{\tilde p^0})^2]\\
    &=\exp[-\frac{\beta}{4\kappa} \int_{t_{\mathcal A_-}}^{t_{\mathcal A_+}} \ud t (-\dv{\vec p}{t}+q \vec E(\vec x,t)+ q \vec v \times \vec B(\vec x,t)+\kappa \frac{ \vec p}{ p^0} )^2].
    \end{aligned}
\end{equation}
Thus, for a pair of any forward and backward trajectories, 
the ratio of the path probabilities between forward and backward trajectories is 
\begin{equation}\label{ft_q0}
    \frac{\Pr[x|x_{\mathcal A_-},p_{\mathcal A_-}]}{\tilde \Pr[\tilde x|\tilde x_{\tilde {\mathcal A}_-},p_{\tilde {\mathcal A}_-}]}=\exp\left[-\beta \int_{t_{\mathcal A_-}}^{t_{\mathcal A_+}} (\dv{\vec p}{t}-q\vec E-q\vec v\times\vec B)\cdot\frac{\vec p}{p^0}\ud t\right]=\exp(-\beta Q^0_{\text{rest}}).
\end{equation}
We have used the expression for the heat in the rest frame 
$Q^0_{\text{rest}}=\int \sum_i(\dv{p^i}{t}-qE^i-q(v\times B)^i)\frac{p^i}{p^0}\ud t$, which can be obtained from Eq. (12) in the main text. 
So far, we have been working in the rest frame. 
The probability ratio in a generic reference frame can be obtained by a Lorentz transformation. 
Note that the Lorentz transformation and the spacetime inversion are exchangeable so that the backward trajectory is independent of the reference frame. 
Furthermore, the change of measure due to the Lorentz transformation cancels out with each other for the forward and the backward pairs. 
Therefore, 
when boosting to other reference frames, the value of the ratio remains the same as $\exp(-\beta Q^0_{\text{rest}})$. 
On the other hand, in a general reference frame, $\exp(-\beta Q^0_{\text{rest}})$ can be re-expressed as $\exp(-\beta u_\mu Q^\mu)$. 
\begin{equation}\label{ft_q}
    \frac{\Pr[x|x_{\mathcal A_-},p_{\mathcal A_-}]}{\tilde \Pr[\tilde x|\tilde x_{\tilde {\mathcal A}_-},\tilde p_{\tilde {\mathcal A}_-}]}=\exp(-\beta u_\mu Q^\mu).
\end{equation}
This is Eq. (14) in the main text. 
The remaining derivation from the above equation to covariant fluctuation theorems is given in Sec. \ref{sec:2}.

\section{work distribution for the driven field}\label{sec:work distribution for the driven field}
In this section, we evaluate the work distribution for a specific driving process of the relativistic scalar field, the first example in the main text. The covariant Jarzynski equality is \added{then }explicitly verified for the work distribution. 
In this process, the higher order potential in Eq. \eqref{c_eom} is set to be $V(\phi)=0$. 
In the infinite past, the external field starts from $h=0$, and it returns to $h=0$ in the infinite future. 
The system starts from equilibrium distribution and finally evolves back to the same equilibrium distribution. 
During the process, the external field performs work, which is then dissipated into the heat bath. 

We proceed by calculating the following generating functional of the field via the path integral approach: 
\begin{equation}
    \begin{aligned}
    \chi[J]&=\ev{\exp(-\int\ud^4xJ(x)\phi(x))}\\
    &=\frac{\int\uD \phi \exp(-\int\ud^4x[\alpha+J(x)\phi(x)])}{\int\uD \phi \exp(-\int\ud^4x\alpha)} .
    \end{aligned}
\end{equation}
Here, $J(x)$ is the counting field. 
Since the process starts at the infinite past, the system thermalizes to equilibrium before the driving field $h$ changes, 
irrespective of the initial distribution in the infinite past. 
Therefore, the integral is over the whole spacetime, and we have ignored the factor of the initial distribution in the above path integral formula. 
Utilizing the Gaussian integral formula, the result is 
\begin{equation}\label{ch}
    \begin{aligned}
    \chi[J]=\exp\{ \int\ud^4x\ud^4y [\frac\kappa\beta J(x)\Delta(x-y)J(y)-J(x)\Delta(x-y)(\partial^\mu\partial_\mu-\kappa^\mu\partial_\mu+m^2)h(y)]\} ,
    \end{aligned}
\end{equation}
where the propagator is defined as 
\begin{equation}
    \Delta(x)=\int\frac{\ud^4p}{(2\pi)^4}\frac{\ue^{\ui p_\mu x^\mu}}{(-p^\mu p_\mu+m^2)^2+(\kappa_\mu p^\mu)^2} .
\end{equation}
The field obeys a Gaussian distribution. 
Noticing the linear relation between the work and the field in Eq. \eqref{wq}, we take the counting field $J$ to be the following form, 
\begin{equation}
    J(x)=-\lambda_\nu\partial^\nu h(x), \quad x\in \mathcal A .
\end{equation}
The generating function for the work 4-vector with the new counting field $\lambda_\nu$ is given by, 
\begin{equation}
    \begin{aligned}
    \chi[-\lambda_\nu\partial^\nu h(x)]=&\ev{\exp(- \lambda_\nu W^\nu)}\\
    =&\exp\{ \int\ud^4x\ud^4y[\frac\kappa\beta \lambda^\mu\partial_\mu h(x)\Delta(x-y)\lambda^\nu\partial_\nu h(y)\\
    &+\lambda^\mu\partial_\mu h(x)\Delta(x-y)(\partial^\mu\partial_\mu-\kappa^\mu\partial_\mu+m^2)h(y)]\} .
    \end{aligned}
\end{equation}
According to the above generating function, the work 4-vector obeys a joint Gaussian distribution, with the following average and covariance, 
\begin{equation}
    \begin{aligned}
    &\ev{W^\mu}=-\int\ud^4x\ud^4y\partial^\mu h(x)\Delta(x-y)(\partial^\nu\partial_\nu-\kappa^\nu\partial_\nu+m^2)h(y),\\
    &\ev{W^\mu W^\nu}-\ev{W^\mu}\ev{W^\nu}= \int\ud^4x\ud^4y\frac{2\kappa}\beta \partial^\mu h(x)\Delta(x-y)\partial^\nu h(y) .
    \end{aligned}
\end{equation}
To verify the relativistic covariant Jarzynski equality for the above generating function, we set $\lambda_\nu=\beta_\nu$ and find
\begin{equation}
    \begin{aligned}
        \ln\ev{\exp(-\beta_\nu W^\nu)}
    &=\int\ud^4x\ud^4y\beta^\nu\partial_\nu h(x)\Delta(x-y)(\partial^\mu\partial_\mu+m^2)h(y)\\
    &=-\int\ud^4x\ud^4y\beta^\nu \partial_\nu(\partial^\mu\partial_\mu+m^2)\Delta(x-y)h(x)h(y) .
    \end{aligned}
\end{equation}
Because $\partial_\nu(\partial^\mu\partial_\mu+m^2)\Delta(x-y)$ is odd under the exchange of $x$ and $y$, 
the above integral vanishes, 
and we have
\begin{equation}
    \ev{\exp(- \beta_\nu W^\nu)}=1 .
\end{equation}
Since the initial and the final states are the same canonical distribution with $h=0$, 
the free energy difference $\Delta F=0$. 
We \added{have }verified the relativistic covariant Jarzynski equality $\ev{\exp(- \beta_\nu W^\nu)}=\exp(-\beta \Delta F)$ in this model.

\section{heat distribution for relativistic Ornstein-Uhlenbeck process}\label{sec:heat distribution for relativistic OU process}
In this section, we  calculate the heat distribution of relativistic Ornstein-Uhlenbeck process, the second example in the main text. 
The equation of motion is nonlinear, making the problem more interesting but hard to solve. 
To simplify, we focus on the 
relaxation process of a $(1+1)$ dimensional ultra-relativistic free particle (\replaced{without}{there is no} external driving force). 
For such an ultra-relativistic particle, the kinetic energy is much larger than the rest energy $\abs{\vec p}\gg m$, leading to the energy-momentum relation $p^0=\abs{ p}$. 
In this (1+1)D case, we use a single letter $p$ to denote the momentum. 
The equation of motion is 
\begin{equation}
    \dv{p}{t}= -\kappa \frac{p}{\abs{p}} +\sqrt{\frac{2\kappa }{\beta}}\xi ,
\end{equation}
where $\xi$ is the standard white noise. 

During a finite time $t$, 
the transition probability from the initial momentum $q$ to the final momentum $p$ can be obtained explicitly (in the rest reference frame of the heat bath), 
\begin{equation}\label{correct}
    \begin{aligned}
    P_t(p|q)=&\frac{1}{\sqrt{4\pi\kappa t/\beta}}\exp[-\frac{(p-q)^2}{4\kappa t/\beta}-\frac \beta 2(\abs{p}-\abs{q})-\frac {\beta\kappa}4]+\frac \beta 4\exp(-\beta \abs{p})\operatorname{erfc}[\frac{1}{\sqrt{4\kappa/\beta}}(\frac{\abs{p}+\abs{q}}{\sqrt t}-\kappa \sqrt{t})] .
    \end{aligned}
\end{equation}
Here $\mathrm{erfc}$ is the complementary error function. 
The above expression is derived by using path integral methods in Ref. \cite{Paraguassu2021}. 
By the way, a typo in the expression of the transition probability in Ref. \cite{Paraguassu2021} is corrected in Eq. \eqref{correct}, where we replace $\sqrt 2$  by $\sqrt{t}$ in the denominator in the last bracket. 

Suppose the initial state of the particle is in thermal equilibrium with the inverse temperature 4-vector $\beta^\mathrm{s}_\mu$, different from the inverse temperature 4-vector $\beta_\mu$ of the heat bath. 
Once being in contact with the heat bath, 
the particle will relax towards the equilibrium at $\beta_\mu$. 
In this process, the system and the bath will exchange both energy and momentum with each other. 
We are interested in the joint distribution of energy-related heat $Q^0=\abs{p}-\abs{q}$ and momentum-related heat $Q^1=p-q$. 
In the rest frame of the heat bath, it can be expressed using the transition probability: 
\begin{equation}
    P(Q^0,Q^1)=\int\ud p \ud q \delta(Q^0-\abs{p}+\abs{q})\delta(Q^1-p+q) P_t(p|q)\mathcal P(q),
\end{equation}
where the initial distribution $\mathcal P(q)$ only depends on the initial momentum $q$. 

\begin{figure}
    \centering
    \includegraphics[width=0.4\textwidth]{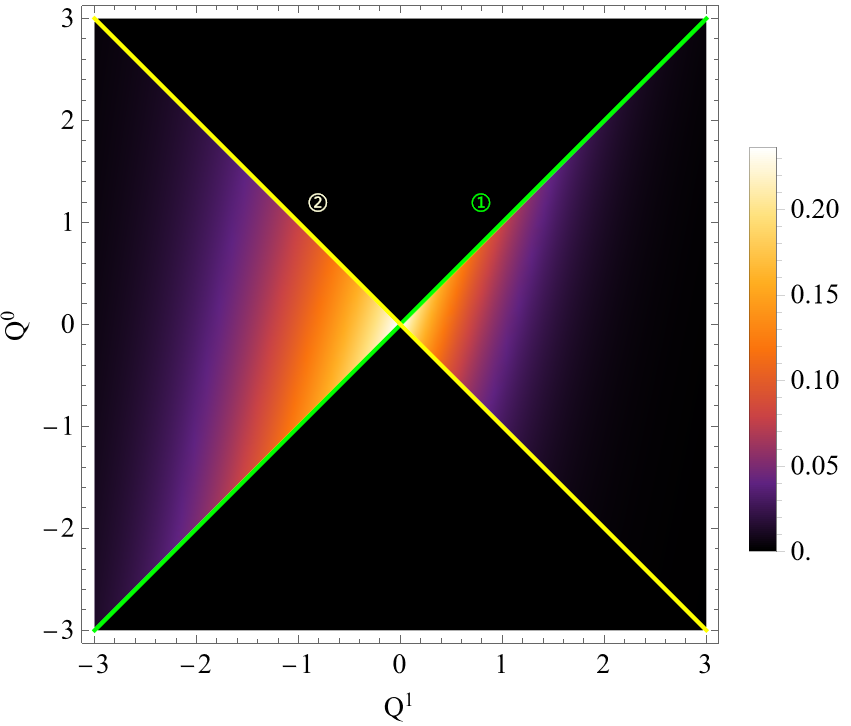}
    \includegraphics[width=0.4\textwidth]{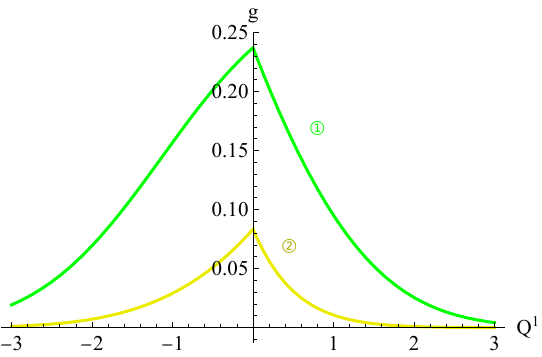}
    \caption{
    Joint distribution of $Q^0$ and $Q^1$ observed in the rest frame of the heat bath: 
    The left panel is the density plot for the joint distribution function. 
    On the two lines $Q^0=\pm Q^1$, distribution is degenerated into $P(Q^0,Q^1)=\delta(Q^0\mp Q^1)g_{\pm}(Q^1)$. 
    In the right panel, function $g_\pm(Q^1)$ is plotted in green and yellow lines for $Q^0=Q^1$ and $Q^0=-Q^1$, respectively. 
    Parameters are chosen as: $\kappa=1$, the initial  inverse temperature 4-vector of the system $\beta^{\mathrm s}_\mu=(5/4,-3/4)$, the inverse temperature 4-vector of the heat bath $\beta_\mu=(1,0)$, and the time duration $t=1$. 
    }\label{heat_distribution}
\end{figure}

After integrating out the delta function, the result of the joint distribution function can be discussed in different situations. 
For $Q^1>0$, 
\begin{equation}\label{27}
    \begin{aligned}
    P(Q^0,Q^1) &= \delta(Q^0-Q^1)\int_0^{+\infty}\ud q P_t(Q^1+q|q)\mathcal P(q)+\delta(Q^0+Q^1)\int_{-\infty}^{-Q^1}\ud q P_t(Q^1+q|q)\mathcal P(q)\\
    &+I_{-Q^1<Q^0<Q^1}P_t\left(\frac{Q^0+Q^1}2|\frac{Q^0-Q^1}2\right)\mathcal P\left(\frac{Q^0+Q^1}2\right).
    \end{aligned}
\end{equation}
Here, $I_{-Q^1<Q^0<Q^1}$ is the indicator function of the region. That is, $I=1$ for $-Q^1<Q^0<Q^1$, $I=0$ otherwise. 
For $Q^1<0$, 
\begin{equation}\label{28}
    \begin{aligned}
    P(Q^0,Q^1)&=\delta(Q^0-Q^1)\int_{-Q^1}^{+\infty}\ud q P_t(Q^1+q|q)\mathcal P(q)+\delta(Q^0+Q^1)\int_{-\infty}^0\ud q P_t(Q^1+q|q)\mathcal P(q)\\
    &+I_{Q^1<Q^0<-Q^1}P_t\left(\frac{-Q^0+Q^1}{2}|\frac{-Q^0-Q^1}{2}\right)\mathcal P\left(\frac{-Q^0-Q^1}{2}\right).
    \end{aligned}
\end{equation}
In both cases, for $ Q^0<\abs{Q^1}$, the distribution is continuous. 
On the $Q^0=Q^1$ and $Q^0=-Q^1$ lines, the distribution is degenerated to a 1d measure. 
For $Q^0>\abs{Q^1}$, the probability is zero. 
This joint distribution is plotted in Fig. \ref{heat_distribution}, where the initial inverse temperature 2-vector of the system is $\beta^{\mathrm s}_\mu=(5/4,-3/4)$, and the inverse temperature 2-vector of the heat bath is $(1,0)$. In this case, the particle initially has the same rest inverse temperature as the bath, $\beta_\mu\beta^\mu=\beta_\mu^{\mathrm s}\beta^{\mathrm s\mu}=1$, but the initial equilibrium state of the particle moves rightwards relative to the bath. 

\replaced{We now examine}{The last thing is about} the heat exchange fluctuation theorem described in the main text. 
The detailed version of the heat exchange fluctuation theorem is 
\begin{equation}\label{29}
    \frac{P(Q^0,Q^1)}{P(-Q^0,-Q^1)}=\exp[-(\beta_\mu-\beta_\mu^\mathrm{s})Q^\mu],
\end{equation}
which leads to an integral version, 
\begin{equation}\label{30}
    \ev{\exp[(\beta_\mu-\beta_\mu^\mathrm{s}) Q^\mu]}=1.
\end{equation}
By choosing a set of parameters and directly computing the joint distribution in Eqs. \eqref{27}-\eqref{28}, it can be checked that Eqs. \eqref{29} and \eqref{30} are indeed satisfied in this model. 

\section{Illuminations on relativistic thermodynamics}\label{formulation}
There have been many different proposals on the  formulation of (traditional macroscopic) relativistic thermodynamics, and researchers are still debating on which formulation is the most appropriate. 
Particularly, there are various viewpoints on  the temperature transformation rule. The temperature of a relativistic moving system is supposed to be lower, higher, invariant, or generalized to an inverse temperature 4-vector. 
However, later studies \cite{Yuen1970,Dunkel2009,K.Nakamura2012} have shown that  these proposals are consistent within their own frameworks (for equilibrium thermodynamics). 
Our study on covariant fluctuation theorems is based on a covariant formulation of relativistic thermodynamics using inverse temperature 4-vector $\beta_\mu$, commonly referred to as the van Kampen formulation or the van Kampen-Israel formulation \cite{Kampen1968,vanKampen1969,Israel1986}.

In this section, we give a brief introduction to the van Kampen formulation of relativistic thermodynamics from the perspective of stochastic thermodynamics. 
\replaced{We also illustrate the main frameworks of other formulations, including the Planck-Einstein and Ott formulations, from the perspective of covariant stochastic thermodynamics. }{We also illustrate how, in the context of equilibrium thermodynamics, existing formulations, such as the Planck-Einstein and Ott formulations, can be derived from the covariant formulation.}
Finally, based on new insights from the stochastic thermodynamics, we remark that the Planck-Einstein formulation and the Ott formulation face severe problems when considering finite-time nonequilibrium thermodynamics. 
In our study, the van Kampen formulation is seamlessly combined with stochastic thermodynamics, making it appropriate for both equilibrium and nonequilibrium contexts. 

\subsection{Basic notions of relativistic distributions}
We provide a pedagogical introduction to relativistic thermodynamics from the perspective of stochastic thermodynamics. 
For simplicity of presentation, we only consider a single-particle (or ideal-gas) system, although similar discussions are valid for field systems as well.

We denote the worldline of the particle as $x^\mu(s)$. 
The kinetic 4-momentum is $p^\mu=m\ud x^\mu/\ud \tau$, with $\tau$ the proper time. The canonical 4-momentum is $P^\mu=(E,\vec P)$. 
We denote by $h(x)$ the external driving parameter which can be tuned during the time evolution, e.g., the volume of the system or external electromagnetic potential. 

We have two typical examples in mind. 
One is the ideal gas inside a box with spacetime region $\mathcal V$. 
This system can be modeled by an infinite square-well potential $U(x)$, with $U(x)=\infty $ outside the box region $\mathcal V$ and $U(x)=0$ inside the box. 
The size of the box is tunable by varying its boundary in a driving process. 
The total energy is $P^0=p^0+U(x)$, and the canonical momentum is equal to the kinetic momentum $P^i=p^i$. 
The other example is a single charged particle under external electromagnetic 4-potential $A^\mu(x)$, which is one of the examples in the main text. 
The canonical 4-momentum is $P^\mu(x,\vec p)=p^\mu+q A^\mu(x)$. 

\begin{figure}
    \centering
    \includegraphics[width=0.5\linewidth]{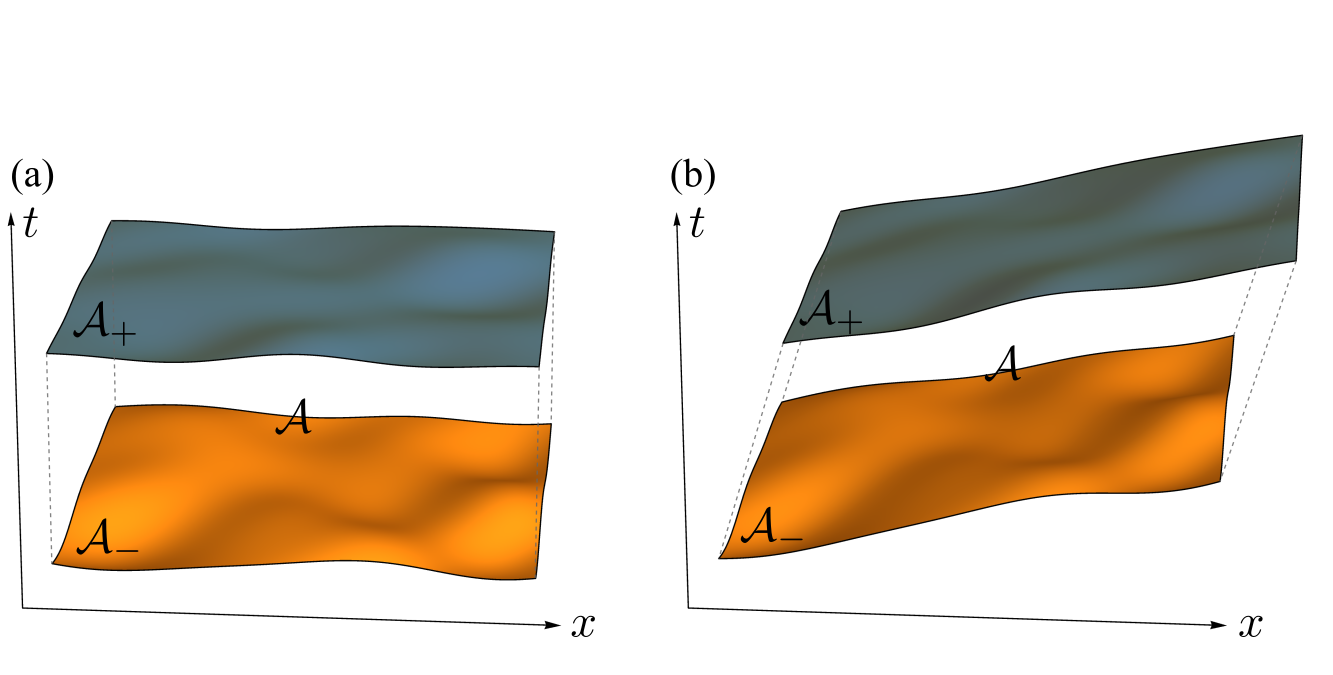}
    \caption{Driving process observed in two different reference frames. (a) Initial and final spacelike hypersurfaces $\mathcal{A}_\mp$ in the lab frame. (b) The same hypersurfaces in a moving frame in which the system (as well as the heat bath) moves rightwards. 
    }
    \label{fig:supple_two_frameworks}
\end{figure}

Due to the relativity of simultaneity, the concept of ``at a specific time" in nonrelativity should be replaced by a more general concept of ``on a spacelike hypersurface" in relativity. Any two spacetime points on such a hypersurface are spacelike (no causal connection). 
One special case of a spacelike hypersurface is a isochronous hyperplane of a certain reference frame. 
To ensure a clear picture of relativistic thermodynamics, this concept of hypersurface is necessary. 
See Fig. \ref{fig:supple_two_frameworks} for an illustration of the initial and final hypersurfaces of a finite-time driving process. 

The distribution function of the single-particle system is denoted as $   \rho(t,\vec x,\vec p) $. 
The ensemble average of an extensive quantity $G$ evaluated on a hypersurface $\mathcal A_0$ is given by 
 \begin{equation}\label{gen_aver}
     \ev{G}_{\mathcal A_0}=\int_{\mathcal A_0}\ud\sigma_\mu \int\frac{\ud^3p}{p^0}p^\mu G(x,\vec p)\rho(x,\vec p) .
 \end{equation}
Here, $\int_{\mathcal A_0}$ denotes the integral on the hypersurface $\mathcal A_0$, and $\ud\sigma_\mu$ is the normal vector of $\mathcal A_0$. 
In the above expression, the kinetic energy $p^0$ should be viewed as a function of $\vec p$ according to the mass-shell condition, $p^0=\sqrt{\vec p^2+m^2}$, and the second integral is only for the 3-momentum. 
We can define a 4-current for $G$ by 
\begin{equation}\label{4-current}
\ev{G^\mu(x)}=\int\frac{\ud^3p}{p^0}p^\mu G(x,\vec p)\rho(x,\vec p).
\end{equation}
The ensemble average of $G$ can be expressed as an integral on the hypersurface, $\ev{G}_{\mathcal A_0}=\int_{\mathcal A_0}\ud\sigma_\mu \ev{G^\mu}$. 
For a isochronous hyperplane $t=t_0$, 
Eq. \eqref{gen_aver} reduces to the following familiar form, 
\begin{equation}
    \ev{G(x,\vec p)}_{t=t_0}=\int\ud^3x\ud^3p G(t_0,\vec x, \vec p)\rho(t_0,\vec x,\vec p) .
\end{equation}
The canonical equilibrium distribution is given by 
\begin{equation}\label{eq_d}
    \rho^{\text{eq}}(t,\vec x,\vec p) = \exp(-\beta_\mu P^\mu(x,\vec p,h))/Z ,
\end{equation}
where $\beta_\mu = \beta u_\mu$ is the inverse temperature 4-vector, with $\beta$ the rest inverse temperature and $u_\mu$ the 4-velocity of the system. 
Normalization factor $Z$ is the partition function. 
According to the formula of equilibrium distribution, two systems are in equilibrium with each other only if all four components of their inverse temperature 4-vectors are equal. 
The derivation of the above equations \eqref{gen_aver}-\eqref{eq_d} can be found in Ref. \cite{Hakim2011}.

Choosing $G$ as the 4-momentum $P^\mu$ in Eq. \eqref{4-current}, we obtain
the energy-momentum tensor, 
with the following expression, 
\begin{equation}
    \ev{T^{\mu\nu}}^\mathrm{eq}= \int\frac{\ud ^3 p}{p^0}p^\mu P^\nu\rho^\mathrm{eq}(x,\vec p) .
\end{equation} 
The total average 4-momentum are calculated by an integral on a specific spacelike hypersurface $ \mathcal A_0$, 
\begin{equation}
    \ev{P^\nu}^\mathrm{eq}_{\mathcal A_0} = \int_{\mathcal A_0}\ud \sigma_\mu \ev{T^{\mu\nu}}^\mathrm{eq} . 
\end{equation}
It should be noted that the average energy and momentum of an equilibrium system depend on the spacelike hypersurface, as we will show in Sec. \ref{app}. 

\subsection{The van Kampen covariant formulation}
Van Kampen introduced the concept of the work 4-vector and heat 4-vector (at ensemble average level) \cite{Kampen1968,vanKampen1969}. 
In the current study, we generalize those concepts from ensemble average quantities to quantities defined on individual stochastic trajectories \cite{Sekimoto2010}. 

For a thermodynamic process between the initial and final hypersurfaces, 
the work 4-vector $W^\mu$ and the heat 4-vector $Q^\mu$ of a stochastic trajectory can be generally defined by Eq. \eqref{wq_par} (see also Eqs. (12) and (13) in the main text). 
Thanks to ideas from stochastic thermodynamics, the physical meaning of $W^\mu$ and $Q^\mu$ is clarified: they are associated with the 4-momentum change due to the external driving, and the 4-momentum exchange between the system and the heat bath, respectively. 
The average of $W^\mu$ and $Q^\mu$ in a quasi-static process is a special case, which we denote as 
$\ev{W^\mu}^\text{qs}$ and $\ev{Q^\mu}^\text{qs}$.

The entropy for an equilibrium system is defined by the following formula, 
\begin{equation}
    \ev S^\text{eq} = \ev{-\ln \rho^{\text{eq}}}^{\text{eq}}_{\mathcal A_0}
\end{equation}
Since $\rho^{\text{eq}}$ is a Lorentz scalar (see Eq. \eqref{eq_d}), $\ev{S}^\text{eq}$ is also a scalar under Lorentz transformation. 
It can be proved that $\ev{S}^\text{eq}$ does not depend on the choice of the hypersurface \cite{Dunkel2009}. 
Therefore, as long as the state of the system does not change, the equilibrium entropy $\ev{S}^\text{eq}$ is invariant.

In the van Kampen formulation, the temperature is generalized to inverse temperature 4-vector $\beta_\mu$.
The covariant second law is given by $\delta\ev{S}-\beta_\mu \delta \ev{Q^\mu}\geq 0$ for a generic thermodynamic process. 
For a quasi-static (reversible) process, the overall entropy production is zero, and we have the following expression
\begin{equation}\label{van}
    \delta \ev S^\text{eq} = \beta_\mu \delta \ev{Q^\mu}^\text{qs} .
\end{equation}
This relation is a covariant generalization of the nonrelativistic relation $T\ud S = \ud Q=\ud E -\mathscr P\ud V$, where $\mathscr P $ is the pressure, and $V$ is the volume. 
 

\subsection{Energy and momentum on different isochronous hyperplanes}\label{app}


The following calculation shows that the average of energy and momentum of an equilibrium system depends on spacelike hypersurfaces (or isochronous hyperplanes). 
In earlier studies, the spacelike hypersurfaces are usually chosen as isochronous hyperplanes, which represents a special case.

We consider two reference frames, the rest frame of the equilibrium system and a moving frame. 
In the moving frame, the system has a velocity $\vec v$ along $z$ direction, $\vec v=(0,0,v)$. 
Physical quantities in the moving frame are denoted by a prime, e.g., $x^\prime$, while quantities in the rest frame are denoted without prime, e.g., $x$.

\begin{figure}
    \centering
    \includegraphics[width=0.2\linewidth]{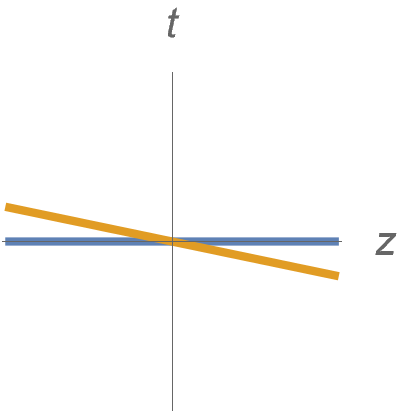}
    \includegraphics[width=0.2\linewidth]{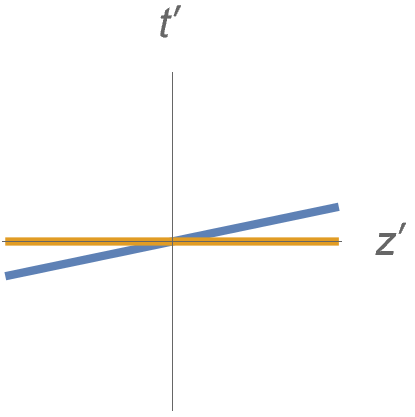}
    \caption{Two isochronous hyperplanes observed in two reference frames: 
    We present two spacetime diagrams, one in the rest frame (left panel) and the other in the moving frame (right panel). 
    Only $t$ and $z$ axes are shown in the diagrams. 
    Two different isochronous hyperplanes, $t=0$ and $t^\prime =0$, are depicted in blue and orange, respectively. 
    The hyperplane $t=0$ is horizontal in the rest frame and is inclined in the moving frame while the hyperplane $t^\prime=0$ is inclined in the rest frame and horizontal in the moving frame. 
    }
    \label{fig:planes}
\end{figure}

We distinguish two different kinds of isochronous hyperplanes, $t=\text{const}$ and $t^\prime=\text{const}$; 
see Fig. \ref{fig:planes}. 
In earlier studies, the isochronous hyperplane in the rest frame is mostly selected to be $t=\text{const}$ while the hyperplane in the moving frame is chosen to be either $t=\text{const}$ or $t^\prime=\text{const}$. 
In the following, we will show how the convention of hyperplane influences the transformation rule of 4-momentum. 

In the rest frame, for an isotropic equilibrium system, the energy-momentum tensor is diagonal, 
\begin{equation}
    \ev{T^{\mu\nu}(x)}^\text{eq}=\operatorname{diag}(\epsilon(\vec x),\psi(\vec x),\psi(\vec x),\psi(\vec x)) .
\end{equation}
$\epsilon(\vec x)$ and $\psi(\vec x)$ only depends on the space but does not depend on time. 
For example, for ideal gas confined in a region $\mathcal V$ with volume $V$ (in the rest frame), 
we have $\epsilon(\vec x)=\epsilon,\psi(\vec x) = \mathscr P$ for $\vec x \in \mathcal V$ and $\epsilon=0,\psi=0$ for $\vec x \notin \mathcal V$, where $\epsilon$ is the energy density, and $\mathscr P$ is the pressure.
For a particle in a confined electromagnetic potential, $\epsilon(\vec x)$ and $\psi(\vec x )$ vary continuously with respect to $\vec x$.

The total 4-momentum on the isochronous hyperplane $t=\text{const}$ is 
\begin{equation}
    \ev{P^{\nu}}^\text{eq}_{t=\text{const}}=\int\ud^3x T^{0\nu}= (\int\ud^3 x \epsilon(\vec x),0,0,0) .
\end{equation}
Only the energy component is nonzero in the rest frame.

Applying the Lorentz transformation, the energy-momentum tensor in the moving frame is 
\begin{equation}
    \ev{T^{\prime\mu\nu}(x^\prime)}^\mathrm{eq} = 
    \begin{pmatrix}
        \gamma^2\epsilon(x)+\gamma^2v^2\psi(x) & 0 & 0& \gamma^2 v \epsilon(x)+\gamma^2 v \psi(x)\\
        0 & \psi(x) & 0 &0\\
        0&0&\psi(x)&0\\
        \gamma^2 v \epsilon(x)+\gamma^2 v \psi(x)&0&0&\gamma^2v^2\epsilon(x)+\gamma^2 \psi(x) .
    \end{pmatrix}
\end{equation}
Here, $\gamma = (1- v^2)^{-1/2}$. 

In the moving frame, there are two choices of isochronous hyperplanes. 
If we evaluate the total 4-momentum on the same hyperplane used in the rest frame, $t=\text{const}$, then 
$\ev{P^{\prime\nu}}_{t=\text{const}}$ is related to $\ev{P^\nu}_{t=\text{const}}$ by Lorentz transformation, 
\begin{equation}
    \ev{P^{\prime0}}_{t = \text{const}} = \gamma \ev{P^0}_{t = \text{const}}, \quad \ev{P^{\prime i}}_{t=\text{const}}= \gamma v^i \ev{P^0}_{t = \text{const}} .
\end{equation}
This Lorentz transformation relation between $\ev{P^{\prime\nu}}_{t=\text{const}}$ and $\ev{P^\nu}_{t=\text{const}}$ also holds for a nonequilibrium state. 
They are the same quantity viewed in different reference frames. 

Next, we consider the total energy-momentum on the isochronous hyperplane\deleted{ of the moving frame} $t^\prime=\text{const}$\added{ in the moving frame}. 
It should be calculated by 
\begin{equation}
    \ev{P^{\prime\nu}}_{t^\prime = \text{const}}^\mathrm{eq} = \int_{t^\prime =\text{const}} \ud^3x^\prime \ev{T^{\prime 0\nu}(x^\prime)}^{\mathrm{eq}} .
\end{equation}
The zeroth component is 
\begin{equation}\label{PE_e}
\begin{aligned}
    \ev{P^{\prime 0}}_{t^\prime = \text{const}}^\mathrm{eq} &= \int_{t^\prime =\text{const}} \ud^3x^\prime[\gamma^2\epsilon(x,y,z)+\gamma^2v^2\psi(x,y,z)] \\
    &=\int \ud^3x^\prime [\gamma^2\epsilon(x^\prime,y^{\prime}, \gamma (z^\prime-vt^\prime_{\text{const}}) )+\gamma^2v^2\psi(x^\prime,y^{\prime}, \gamma (z^\prime-vt^\prime_{\text{const}}))]\\
    &= \int\ud^3 x^\prime [\gamma\epsilon(\vec x^\prime)+\gamma v^2 \psi(\vec x^\prime )] = \gamma \ev{P^{ 0}}_{t=\text{const}}^\mathrm{eq} +\gamma v^2 \int\ud^3x\psi(\vec x). 
    \end{aligned}
\end{equation}
In the second equality, $\vec x(t^\prime,\vec x^\prime)$ is substituted by $t^\prime,\vec x^\prime$. 
In the third line, we applied a change of variable $z\rightarrow{z^\prime}$, giving a factor $\gamma^{-1}$ (can be viewed as the effect of volume change). 
Similarly, the momentum is 
\begin{equation}\label{PE_p}
    \ev{P^{\prime 3}}_{t^\prime = \text{const}}^\mathrm{eq}=\gamma v \ev{P^0}_{t=\text{const}}^\mathrm{eq}+\gamma v \int\ud^3 x \psi(\vec x).
\end{equation}
$\ev{P^{ \nu}}_{t = \text{const}}$ and $\ev{P^{\prime \nu}}_{t^\prime = \text{const}}$ are not related to each other by Lorentz transformation. 
For ideal-gas system, the quantity $\int\ud^3 x \psi(\vec x)=\mathscr P V$ in the additional term is the pressure times the rest volume. 
Note that the transformation rule Eq. \eqref{PE_e} and \eqref{PE_p} is valid only for equilibrium system. 
For a generic nonequilibrium state, no transformation rule of two 4-momenta can be given because they are evaluated on different spacetime points. 

Different formulations adopt different conventions of the isochronous hyperplane. 
In the Planck-Einstein formulation, the hyperplane is chosen as $t^\prime=\text{const}$ in the moving frame, different from that in the rest frame. 
In the Ott and van Kampen formulations, the hyperplane is fixed to be $t=\text{const}$, independent of the frame. Furthermore, in the Ott and van Kampen formulations, the hyperplane can be generalized to a generic fixed spacelike hypersurface.

To summarize, in the moving frame, two kinds of 4-momentum on different isochronous hyperplanes, $\ev{P^{\prime\nu}}_{t^\prime=\text{const}}^\mathrm{eq}$ and $\ev{P^{\prime\nu}}_{t=\text{const}}^\mathrm{eq}$ are different. 
As we will see later, the former choice of the hyperplane (used in the Planck-Einstein formulation) is problematic when considering nonequilibrium processes.


\subsection{Existing formulations of relativistic thermodynamics\deleted{ derived from covariant formulation}}
Three formulations are commonly discussed in relativistic thermodynamics \cite{Yuen1970,Dunkel2009,K.Nakamura2012}, the Planck-Einsten formulation, the Ott formulation, and the van Kampen formulation. 
In this subsection, we briefly illustrate the main frameworks of the three formulations, in the context of equilibrium thermodynamics. 
Also, based on new insights from stochastic thermodynamics, we provide additional remarks on these formulations, especially regarding irreversible thermodynamic processes. 

\subsubsection{Setup}

Earlier works on relativistic thermodynamics mainly consider equilibrium thermodynamics and reversible (quasi-static) processes. 
Instead, we will consider both equilibrium and nonequilibrium processes. 

The discussion for equilibrium thermodynamics partially follows Ref. \cite{Dunkel2009}, which points out that two key points account for the difference  of existing formulations: 
choices of isochronous hyperplane, and definitions of heat. 
First, different choices of isochronous hyperplane lead to different energy and momentum transformation rules. 
Second, 
since all formulations admit that the entropy is a scalar under Lorentz transformation, 
the different definitions of heat directly lead to the discrepancy of the temperature transformation rule.

We first specify the setup for the equilibrium thermodynamics. 
We suppose the equilibrium thermodynamic system is at rest in the lab frame (\replaced{also}{or} referred to as rest frame). 
Meanwhile, we consider a moving frame, in which the system is moving at 3-velocity $\vec v=(0,0,v)$ along $z$-direction. 
The corresponding 4-velocity is $u^\mu=(\gamma,\gamma \vec v)$ with $\gamma=(1-v^2)^{-1/2}$. 
The quantities in this moving frame is denoted by a prime, e.g., $t^\prime$. 
In this section, we still use $W^\mu$, $Q^\mu$, and $\beta_\mu$ to denote work, heat, and inverse temperature 4-vector in covariant stochastic thermodynamics. 
The thermodynamic quantities in existing formulations are denoted differently by corresponding subscripts. 

An important assumption in most existing formulations is that in the lab frame, the spatial components of work and heat vanishes, $\ev{W^i}=0$, $\ev{Q^i}=0$. 
Actually, this assumption is generally violated in a finite-time driving process. 
Even if the both initial and final states are at rest, the external driving can influence the momentum by $\ev{W^i}$, which is compensated by the momentum exchange with the heat bath, $\ev{Q^i}=-\ev{W^i}$. 
We will come back to this point in the discussion later.

\subsubsection{Planck-Einstein formulation}\label{Planck-Einstein}
In the Planck-Einstein formulation, moving objects get cooler. 
We denote the quantities in this formulation with a subscript ``cool". 
When evaluating energy and momentum in the moving frame, the isochronous hyperplane is chosen to be  $t^\prime=\text{const}$, which is different from the isochronous hyperplane in the lab frame $t=\text{const}$. 
In terms of covariant 4-vectors, the energy and momentum in the moving frame are given by 
$E_\text{cool}^\prime=\ev{P^{\prime 0}}^\mathrm{eq}_{t^\prime =\mathrm{const}}$ and 
$\vec P_\text{cool}^\prime = \ev{\vec P^\prime}^\mathrm{eq}_{t^\prime=\text{const}}$.

According to Eqs. \eqref{PE_e} and \eqref{PE_p}, 
the energy and the momentum transform as 
\begin{equation}\label{PE_er}
 E_\text{cool}^\prime= \gamma E_\text{cool} +\gamma v^2 \int\ud^3x\psi(\vec x), \quad 
   \vec P_\text{cool}^\prime = \gamma\vec v E_\text{cool}+\gamma\vec v \int\ud^3 x \psi(\vec x).
\end{equation}
For ideal gas inside a box, we have $\int\ud^3 x \psi(\vec x)=\mathscr P V$. 
Since the isochronous hyperplanes are chosen differently, the energy and the momentum do not transform as a Lorentz vector. 

In most earlier studies, the change of volume is usually considered as the only way to perform work. 
In the Planck-Einstein formulation, this volume work in the moving frame is defined by a quasi-static expression for reversible processes, 
\begin{equation}\label{w_c}
    \delta W_\text{cool}^\prime=\mathscr P \delta V^\prime =\gamma^{-1} \mathscr P \delta V =\gamma^{-1} \delta W_\text{cool}. 
\end{equation}
The second equality uses the relativistic volume transformation rule, $V^\prime=\gamma^{-1} V^\prime$, and the pressure $\mathscr P$ is invariant.

In the Planck-Einstein formulation, heat in the moving frame is defined as
\begin{equation}\label{q_c}
     Q_{\text{cool}}^\prime = \Delta E_\text{cool}^\prime  -  \vec v \cdot\Delta \vec P^{\prime }_\text{cool} -W^\prime_\text{cool} .
\end{equation}
In the above definition, the quantity related to momentum, $\vec v\cdot \Delta\vec P_\text{cool}$, is subtracted from the energy change. 
According to Eq. \eqref{PE_er}, we have $ E_\text{cool}^\prime  -\vec  v\cdot\vec  P^{\prime }_\text{cool}=\gamma^{-1} E_\text{cool}$, and additional terms cancel out in this combination. 
Therefore, the heat in the moving frame is related to that in the rest frame by 
\begin{equation}
    Q_{\text{cool}}^\prime=\gamma^{-1} Q_\text{cool} .
\end{equation}

The temperature is defined via the thermodynamic relation $T_\text{cool}\delta S^\mathrm{eq}= \delta Q_{\text{cool}}$. 
Since the entropy change is invariant, the temperature is proportional to the heat.
Therefore, in moving frame, the temperature becomes lower, $T_\text{cool}^\prime=\gamma^{-1}T$, where $T$ is the rest temperature.

There are conceptual disadvantages in the definitions of the work and heat by Eqs. \eqref{w_c} and \eqref{q_c}. 
The expression of the quasi-static volume work \eqref{w_c} is not justified for a moving volume, and we observe that it cannot be directly related to the work 4-vector $W^\mu$ defined from stochastic thermodynamics. 
In addition, in Eq. \eqref{q_c}, the quantity $E^\prime-\vec v\cdot \vec P^{\prime }$ cannot be regarded as the internal energy since it is different from the energy in the mass-center frame. 
Therefore, the physical meaning of $Q_{\text{cool}}$ defined in Eq. \eqref{q_c} is unclear, which is a drawback of this formulation. 
In addition, this formulation does not provide a covariant framework, which is another disadvantage.
Nevertheless, once adopting those definitions, the theoretical framework of the Planck-Einstein formulation is self-consistent in the context of equilibrium thermodynamics. 

A more severe problem occurs in the Planck-Einstein formulation when considering nonequilibrium thermodynamics. 
This problem is  related to the choice of isochronous hyperplanes. 
The transformation rule of energy and momentum, Eq. \eqref{PE_er}, which already differs from Lorentz transformation, is based on the equilibrium expression of energy-momentum tensor. 
For a general nonequilibrium time-dependent distribution, 
the energy and momentum in the rest frame and in the moving frame can be completely irrelevant since they are evaluated on two different hyperplanes. 

\begin{figure}
    \centering
    \includegraphics[width=0.3\linewidth]{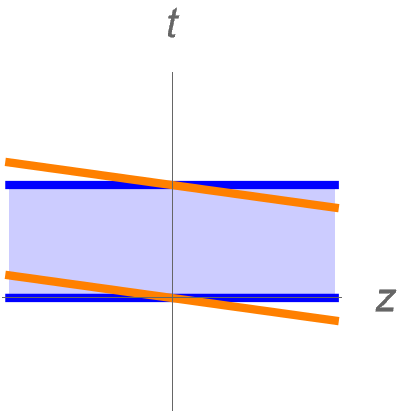}
    \caption{A finite-time process takes place in the spacetime region $t_0<t<t_1$: the initial and finial hyperplanes are depicted in blue. 
    The light blue shadow denotes the spacetime region of the driving process. 
    Orange lines show the isochronous hyperplanes chosen in the Planck-Einstein formulation in the moving frame, $t^\prime=\text{const}$, and they do not coincide with the blue hyperplanes. }
    \label{filling}
\end{figure}

The situation gets even worse in a finite-time driving process. 
Recall that a finite-time process starts and ends on two specific spacelike hypersurfaces (or hyperplanes), which must remain unchanged in different reference frames. 
For example, we suppose the driving process starts at the hyperplane $t=t_0$ and ends at the hyperplane $t=t_1$, corresponding to two isochronous hyperplanes in the rest frame. 
In other words, the driving process happens in the spacetime region $t_0<t<t_1$. 
See Fig. \ref{filling} for an illustration. 
However, in the moving frame, energy and momentum in the Planck-Einstein formulation are evaluated on another set of hyperplanes, $t^\prime =\text{const}$, which cannot be related to the initial and final hyperplanes of the driving process, $t=t_0,t_1$. 
With Planck-Einstein's convention of the isochronous hyperplane, this finite-time process cannot be properly addressed in the moving frame. 
Hence, it is hard to extend the Planck-Einstein formulation to nonequilibrium regime. 


\subsubsection{Ott formulation}\label{Ott}
In the Ott formulation, moving objects get warmer. We denote the quantities in this formulation with a subscript ``warm".
The isochronous hyperplane is fixed to be $t=\text{const}$ for every frame (this hyperplane can be generalized to a fixed spacelike hypersurface). 
The energy and momentum in the moving frame is given by $P^{\prime\mu}_\text{warm}=\ev{P^{\prime\mu}}_{t=\text{const}}$, 
which transforms as a Lorentz vector. 

Heat in the moving frame is defined as the energy-related component of the heat 4-vector $Q^\mu$, 
\begin{equation}\label{q_h}
    Q_{\text{warm}}^\prime=\Delta \ev{P^{\prime 0}}- \ev{W^{\prime 0}}=\ev{Q^{\prime 0}}.
\end{equation}
Assuming $\ev{Q^i}=0$ in the rest frame, we have the following relation between heat in the lab frame and the moving frame, 
\begin{equation}\label{q_h_r}
    Q_{\text{warm}}^\prime=\gamma \ev{Q^0}-\gamma \vec v\cdot\ev{\vec Q}=\gamma \ev{Q^0} .
\end{equation}
The temperature is defined by the thermodynamic relation $T_\text{warm}\Delta S^\mathrm{eq}= \Delta Q_{\text{warm}}^\text{qs}$. 
Since $\Delta S^\mathrm{eq}$ is invariant, the temperature in a moving frame becomes warmer, 
$    T^\prime_\text{warm} = \gamma T $, where $T$ is the rest temperature.

In the Ott formulation, there is an obvious disadvantage in the definition of heat and work: the momentum-related components are not taken into account. In addition, this definition leads to the  violation of the principle of covariance. Nevertheless, once adopting those definitions, the framework is self-consistent in equilibrium regime. 


When considering nonequilibrium thermodynamics, a more severe drawback is that the Clausius inequality in the moving frame, 
$\Delta S\geq Q^{\prime}_{\text{warm}} /T^{\prime}_{\text{warm}}$, could be violated in generic irreversible processes. 
Note that the derivation of Eq. \eqref{q_h_r} uses the assumption that no momentum is exchanged between the system and the heat bath in lab frame, $\ev{Q^i}=0$. 
However, this assumption is usually violated in a finite-time process. 
Even if the initial and final states of the system are both at rest in the lab frame, the external driving can influence the system momentum by  $\ev{W^i}$ during the process, which is compensated by the momentum exchange with the heat bath $\ev{Q^i}=-\ev{W^i}$, so that the velocity change between initial and final states vanishes, $\Delta \ev{P^i}=0$. 
For example, this happens in a finite-time gas compression process if the volume changes asymmetrically. 
The moving boundary influences the momentum of the gas molecules by collision, and the extra momentum from the moving boundary is finally released into the bath. 
In this case, the second equality in Eq. \eqref{q_h_r} is invalid, i.e., $Q_{\text{warm}}^\prime\neq\gamma \ev{Q^0}$. 
Consequently, the Clausius theorem from Ott's definition of heat and temperature, $\Delta S^\prime\geq Q_{\text{warm}}^\prime/T_{\text{warm}}^\prime$ can be violated. 

This severe problem can be remedied by changing the definition of heat to take into account momentum-related components of heat, which will lead to the van Kampen formulation. 


\subsubsection{Van Kampen formulation}
In the van Kampen formulation, 
the covariant thermodynamic relation is given by (Boltzmann constant is set to $1$)
\begin{equation}
    \Delta S=\beta_\mu \ev{Q^\mu}^\text{qs}. 
\end{equation}
Another way to write this relation is 
\begin{equation}
    \Delta S=\frac{u_\mu}{T} \ev{Q^\mu}^\text{qs} .
\end{equation}
Here, $T$ is what we call the rest temperature. 
The relation between $T$ and the inverse temperature 4-vector is 
$\beta_\mu=u_\mu/T$ and $T=(\beta_\mu\beta^\mu)^{-1/2}$. 

Therefore, in the van Kampen formulation, there are two concepts of temperature, the rest temperature and the inverse temperature 4-vector. 
Some researchers, such as Landsberg \cite{LANDSBERG1966}, regard the rest temperature as temperature.
This is just a preference of terminology. 
However, to avoid any confusion, we prefer using ``rest temperature" for $T$ and using ``inverse temperature 4-vector" for $\beta_\mu$. 

The van Kampen formulation can be seamlessly combined with stochastic thermodynamics, and it is suitable to describe a generic irreversible thermodynamics process. 
One evidence is the promotion of fluctuation theorems into covariant forms. 



\subsection{Remarks}

Studies of relativistic thermodynamics have previously focused on equilibrium thermodynamics, and nonequilibrium relativistic thermodynamics remains largely unexplored. 
Stochastic thermodynamics brings new insights by exploring irreversible thermodynamic processes. Based on these new insights, we distinguish equilibrium and nonequilibrium regarding the controversy over relativistic thermodynamic theories:

\begin{itemize}
    \item \textbf{Equilibrium thermodynamics:}
In the Planck-Einstein and Ott formulations, there are disadvantages regarding the lack of covariance and the physical meaning in the definitions of heat, work, and temperature. There is no such problems in the van Kampen formulation. 
Nevertheless, once accepting their definitions, all three frameworks (Planck-Einstein, Ott, and van Kampen) are self-consistent. 

\item \textbf{Nonequilibrium thermodynamics:}
The Planck-Einstein and Ott formulations face severe problems in dealing with irreversible processes; see the discussions in Sec. \ref{Planck-Einstein} and \ref{Ott}. Conversely, the van Kampen formulation integrates seamlessly with stochastic thermodynamics, offering a natural framework for relativistic thermodynamics of irreversible processes.
\end{itemize}





Therefore, the van Kampen formulation stands out among three formulations. 
Hopefully, a consistent theory of relativistic thermodynamics, including nonequilibrium contexts, will be built based on the van Kampen formulation in the future. 

\bibliography{covariant_title_protected}